\documentclass[preprint,preprintnumbers,amsmath,amssymb]{revtex4-1}
\usepackage{txfonts}
\usepackage{amsfonts}
\usepackage{amsmath}
\usepackage{graphicx} 
\usepackage{dcolumn}  
\usepackage{bm}
\usepackage{overpic}
\usepackage{color}
\begin{document}

\title{Conditions for Viral Influence Spreading through Multiplex Correlated Social Networks}

\author{Yanqing Hu$^{1}$, Shlomo Havlin$^2$,
  Hern\'an A. Makse$^1$}

\affiliation{ 
$^1$ Levich Institute and Physics Department, City College of New
  York, New York, New York 10031, USA \\$^2$Minerva Center and Physics
  Department, Bar-Ilan University, Ramat Gan 52900, Israel}

\begin{abstract}

A fundamental problem in network science is to predict how certain
individuals are able to initiate new networks to spring up ``new
ideas''. Frequently, these changes in trends are triggered by a few
innovators who rapidly impose their ideas through ``viral'' influence
spreading producing cascades of followers fragmenting an old network
to create a new one. Typical examples include the raise of scientific
ideas or abrupt changes in social media, like the raise of
Facebook.com to the detriment of Myspace.com.  How this process arises
in practice has not been conclusively demonstrated. Here, we show that
a condition for sustaining a viral spreading process is the existence
of a multiplex correlated graph with hidden ``influence
links''. Analytical solutions predict percolation phase transitions,
either abrupt or continuous, where networks are disintegrated through
viral cascades of followers as in empirical data. Our modeling
predicts the strict conditions to sustain a large viral spreading via
a scaling form of the local correlation function between multilayers,
which we also confirm empirically.  Ultimately, the theory predicts
the conditions for viral cascading in a large class of multiplex
networks ranging from social to financial systems and markets.

\end{abstract}



\maketitle

\section{Introduction}

People's adoption of new ideas or products and even novel scientific
theories often depends upon the foregoing support of a few innovators,
pioneers or knowledgeable individuals. These early adopters
disseminate the new idea through viral influence spreading that leads
to cascades of followers
\cite{ThomasSchelling,SantosRev,caldarelli,Newman-book,ThomasValente,EverettRogers}. Examples
are found in the rise of brand-new consumer products, where a few
early-adopters can have a large effect in the entire population, a
process leading to modern engagement strategies of ``viral
marketing''. Similarly, scientists in a given academic field work in
specialized topics by developing collaboration networks.  As
innovative ideas arise, the majority of them may rapidly transition to
the study of the new topic.  When the pioneers migrate to develop a
new idea, their former social network weakens its roots
hence, leading to its rapid disintegration
\cite{MarkGranovetter,WattsCascading,KleinbergInfluence1,sergey,korniss}.
A prominent example is the rise of the social networking community
Facebook.com to the detriment of the previously dominant Myspace.com,
as shown in Fig.  \ref{PercolationNetwork}a. The conditions favoring
such a process of disintegration at the tipping point of a dominant
network or community and the raise of a competing network have not
been conclusively demonstrated so far.


Typically, the innovators exert their influence not only through the
regular channels of communication in their social networks--- such as
mutual connectivity through friendship, collaborations, family or
other types of direct contact--- but through unidirectional links
based upon their ``cognition influence''. This means that scientific
innovators, for instance, would lead the introduction of new
scientific ideas by engaging learners
through ``links of influence''.  These are hidden directed links that
can be found in many situations, e.g., people following trends set by
popular singers and actors,
even though the actors do not ``know'' their followers. Equally
important are financial networks and markets, where one company or
bank may depend on others due to financial or technical reasons
\cite{RoniPNAS}. This illustrates a fundamental property of social
networks: while network functionality depends on a layer of mutual
connectivity links,
the network stability depends on a hidden ``guided influence'' network
quantified by the state of being influenced by knowledgeable
individuals.

Systems consisting of network layers with multiple types of links
  such as those treated here are referred to as multiplex networks
  \cite{m1,m2,m3}, that, in some cases, are equivalent to
  interdependent networks \cite{sergey}.  Such a network of networks
  structure is shown to be crucial for cascading failure
  \cite{sergey}, transport \cite{51}, diffusion \cite{52}, evolution
  of cooperation \cite{perc}, competitive percolation \cite{50} and
  neuronal synchronization \cite{53}.  Specifically related to
  spreading processes, previous research has addressed the spreading
  of human cooperation in multiplex networks \cite{perc}, showing that
  it depends significantly on the properties of the correlations
  between network layers, as described in \cite{perc2}. Our approach
  has close connection with recent work on generalized percolation
  \cite{boguna}. Furthermore, the percolation modeling that we apply
  is related to the study of percolation in multiplex networks in the
  context of interdependent networks as studied in \cite{baxter,son}.


\section{Results}

The multiplex structure can be investigated in the collaboration
networks formed by scientists \cite{newman-colla} and
in online social blogging communities of information dissemination
such as LiveJournal.com \cite{lev,livejournal}.  Formally, we consider a
network with two types of links: connectivity and influence links. The
{\bf connectivity links} are undirected and correspond to a close
relationship between two nodes: for instance, scientists who have
coauthored at least $s$ articles during a specified interval in the
journals of the American Physical Society (APS). Underlying this basic
structure, there is a hidden network of {\bf influence links} among
scientists that can be quantified through the citation list of
papers. When author A systematically cites papers of author B, then we
assume that there is a directed outgoing influence link A $\to$ B.
Similar structure can be defined in the LiveJournal network of
information dissemination, which will be studied below.


We start by tracking the upsurge and dissaperance of Physics trends
through the creation and removal of fields in the Physics and
Astronomy Classification Scheme (PACS) compiled by the APS from 1975
to 2010 (see Supplementary Information for details).
 PACS is a hierarchical classification of the literature in the
 physical sciences where each published paper is assigned one or more
 PACS numbers. Figure \ref{PercolationNetwork}b shows the number of
 scientists in the largest connected component (analogous to the giant
 component in the thermodynamic limit) of scientists in the
 Statistical Physics community which, until around 1998, was
 publishing under PACS 64.60: ``General Studies of Phase
 Transitions''. After 2000, many of those researchers quickly switched
 to publish in the field of ``Complex Networks'' (under a series of
 PACS 89.75-k created in 2001). The number of scientists in the
 giant component of the collaboration networks
 (Fig. \ref{PercolationNetwork}b) shows a similar behavior as myspace
 and facebook users in Fig. \ref{PercolationNetwork}a. This similarity
 hints at possible generic features acting when a new trend competes
 with an older trend for the same pool of users.

To quantify the viral cascading process, we perform a real-time
percolation analysis \cite{Callaway,cohena} on the most important
fields contributing to the rise of Complex Networks. These fields
include: Fluctuations (PACS 05.40), Chaos (PACS 05.45), Phase
Transitions (PACS 64.60), Thermodynamics (PACS 05.70), and Interfaces
(PACS 68.35). These fields have experienced either a decline or slower
growth in the number of published papers after 2000 as shown in
Fig. \ref{PercolationNetwork}c.  We notice that scientists publishing
in Complex Networks still include in their papers the original PACS
number of, for instance, Phase Transitions or Chaos.  Therefore, the
decay in the frequency of citations plotted in Fig
\ref{PercolationNetwork}c is not very large, even though scientists
are already working in the field of Complex Networks. This situation
explains why the frequency of citation in Fig.
\ref{PercolationNetwork}c does not decay as rapidly as the giant
component in Fig. \ref{PercolationNetwork}b, but remains relatively
constant or present a small decay after scientists start to publish in
Complex Networks.

\subsection{Empirical results on APS collaboration networks}

We calculate the size, $N_\infty$, of the giant connected component of
authors in each field from the articles published in the pre-Complex
Network period (1997--2001).  Then, we identify the fraction $1-p$ of
pioneers from each field defined as those scientists who published at
least one paper in Complex Networks in 2001. To quantify the effect of
cascades we measure the size of the largest component of the original
collaboration network, $n_\infty$, at a later time (2005-2009). Figure
\ref{important_figure}a shows the fast decay of the fraction of nodes
in the largest connected component which we call
$P_\infty=n_\infty/N_\infty$ with $1-p$ for the different PACS fields
as evidenced of the rapid disintegration of each physics community.
We notice that the fraction of departing pioneers is $1-p$. Thus, Fig.
\ref{important_figure}a should be interpreted, for instance for the
field of Phase Transitions, as follows: a small fraction $1-p=10$\% of
departing pioneers leads to a large 81\% shrinking in the largest
component of the network. The cascading behavior triggered by the
departure of $1-p$ pioneers is visualized in the influence network
representation of Fig. \ref{important_figure}b and the connectivity
representation in Fig. \ref{important_figure}c (see also SI
Fig. \ref{PioF1Other}).

 In addition of considering the largest component, many networks
 consist of other clusters. For this reason, we have also calculated
 the size of the second largest clusters for the five networks
 involved in the APS communities: Chaos, Fluctuations, Interface,
 Phase Transitions and Thermodynamics. We find that the size of the
 second largest clusters are 31, 21, 49, 29, 46, respectively. These
 numbers are small compared with the size of the largest
 components of the networks: 1126, 522, 232, 193, 87, respectively. In
 principle, the second largest clusters can be analyzed in the same
 way as the largest component. However, we find that the size of
 the second largest clusters are too small to obtain meaningful
 statistical results.

The disintegration process can be interpreted as a percolation phase
transition at a critical threshold $p_c$, defined when
$P_{\infty}(p_c)= 0$. For each network, $1-p_c$ quantifies the minimum
fraction of departing nodes (pioneers) who are able to breakdown the
network \cite{Callaway,cohena}. The data seems to percolate at
remarkably large values of $p_c$
(Fig. \ref{important_figure}a) implying that the collaboration
networks are highly vulnerable. This result is in sharp contrast to
the prediction of classical percolation theory on scale-free networks
without influence links: $p_c=0$
\cite{Callaway,cohena,barabasi-attack,satorras01,NewmanSpread}
(collaboration networks have been found to have a power-law tail in
the degree distribution \cite{newman-colla}, $P(k)\sim
k^{-\gamma}$ with $\gamma < 3$, see also SI). The prediction $p_c=0$
exemplifies the extreme resilience of scale-free networks under random
removal of nodes.

In principle, a plausible explanation of the extreme fragility of the
scientific communities could be that the respective networks are being
disrupted by the departure of the most connected people: scale-free
networks are resilient to random removal of nodes ($p_c=0$)
\cite{cohena,satorras01}, but they are very vulnerable ($p_c$ close to
1) in regard to hub departure \cite{barabasi-attack}. However, we find
empirically that the pioneers are not the highly connected
scientists. Yet, they are minor players who develop a novel appealing
idea leading to the creation of an entire new community and to the
disintegration of the old system (see Fig. \ref{important_figure}d and
SI Table \ref{table1}).
This collapse is particularly true since the most well-connected
individuals follow the new trend sustaining a viral cascade of
influences.

We compare the average connectivity degree of pioneers in the
  largest connected component $\langle k_{\rm pio}\rangle$ for
  each PACS community and found that $\langle k_{\rm pio}\rangle$ is
  much smaller than the maximum degree in the network, $k_{\max}$, as
  shown in Fig.  \ref{important_figure}d and SI Table \ref{table1}.
  Furthermore, $\langle k_{\rm pio}\rangle$ is smaller or of the same
  order than the average degree of the nodes, $\langle k \rangle$.
  This result indicates that the pioneers are not always the hubs in
  the network. Instead, the pioneers have a degree which is very close
  to the randomly selected nodes.



\subsection{Percolation modeling}

To identify the conditions for viral cascading leading to network
fragility, we develop a generic model and search the space of
solutions by calculating the percolation threshold. The network
contains undirected connectivity links and directed influence links.
Each node in the network is characterized by the degree $k$ of its
connectivity links, the degree $k_{\rm in}$ of incoming influence
links, and the degree $k_{\rm out}$ of outgoing influence links
(Fig. \ref{model}a, by definition $\langle k_{\rm in} \rangle =
\langle k_{\rm out} \rangle$). In the most general case, these
quantities are correlated as measured by the joint probability
distribution function: $P(k,k_{\rm in},k_{\rm out})$. Indeed, below we
demonstrate that viral cascades can be sustained only when there is a
positive correlation between $k$ and $k_{\rm out}$, which
indeed we find empirically (Fig. \ref{model}g and Fig. \ref{lj}b).




We demonstrate a cascading process (Fig.~\ref{model}a-f) initiated by
removal of a node who creates a new idea and moves to a new field of
science. We map out this process to a correlated percolation model to
find analytical solutions to predict $p_c$, as well as the universal
boundaries of the phase diagram over an ensemble of correlated random
graphs. The main analytical treatment of the problem is based on the
method of generating functions \cite{Callaway,PerLong,sergey}. We
generalize the previous uncorrelated theory
\cite{Callaway,PerLong,sergey} to the case of a correlated network
using:

\begin{equation}
G(x_1,x_2,x_3)=\sum_k\sum_{k_{\rm in}}\sum_{k_{\rm out}}
P(k,k_{\rm in},k_{\rm out}) x_1^kx_2^{k_{\rm in}}x_3^{k_{\rm out}}.
\end{equation}


At the heart of the model, there is a cascading process mimicking the
departure of nodes following influential nodes. Such a process is
described by a certain probability $q_h$ which is estimated from the
data and determines the departing process as follows.  Imagine that
node A is following $k_{\rm out} = 15$ other nodes. The model takes
into account that node A will leave the network with a certain
probability $q_h$ when one of his 15 influential nodes leave. This
probability is a parameter of the model and is determined from the
experimental data as analyzed in Fig. \ref{akin} and Section
\ref{influence_links}. For the sake of argument, imagine that node A
will follow departing nodes with low probability, let's say
$q_h=0.2$. Implementing such a probability per each link of node A
would imply that node A would have a 20\% chances to leave the network
when one of the $k_{\rm out} =15$ followees departs. A direct
implementation of this rule would lead to a rather intractable model
from the mathematical point of view. Rather than implementing this
probability into the model directly, we perform a mapping to a
completely equivalent process: we first create an equivalent network
where we reduce the number of original out-going links $k_{\rm out}$
links to an equivalent $k^{\rm equiv}_{\rm out}$, for instance, for
node A from 15 to 3 (that is, $k^{\rm equiv}_{\rm out} = q_h \times
k_{\rm out} = 3$). We then consider that if any of the 3 nodes in the
equivalent network leave, then node A leaves with probability one. In
the statistical ensemble, both networks are fully equivalent. The main
parameter of the model is the average effective out-going links
$\langle k_{out} \rangle$, which is obtained from the real data as
explained in Fig. \ref{akin} and Section \ref{influence_links}. This
mathematical trick, which has been introduced previously in
\cite{NewmanSpread}, renders an untractable mathematical model,
tractable.  The probability $q_h$ which determines the effective
out-going links is a parameter of the model and we will show in
Fig. \ref{important_figure} that the experimental data on the five
considered APS communities are within the upper and lower bound
predicted by the theory of the effective $\langle k^{\rm equiv}_{\rm
  out} \rangle = 0.44$ and $\langle k^{\rm equiv}_{\rm out} \rangle =
0.83$.

Figure \ref{model}a-f illustrates the cascading process in a
  simple network considered as the giant component in the model (black
  links, red nodes) plus the influence directed network (green
  links). For simplicity we describe the process in the equivalent
  network, which is the one that is solved analytically. In
  Fig. \ref{model}a, a given pioneer departs as indicated.  Such a
  departure produces a regular percolation process (not cascading) of
  disconnecting two other nodes from the giant component as seen in
  Fig. \ref{model}b. Additionally, the pioneer node produces an
  influence-induced departure of an extra node as indicated in
  Fig. \ref{model}c, which in turns produces another two
  influenced-induced departures as shown in the same figure. At this
  point, the cascading starts since the influence-induced departures
  produce extra percolation disconnections from the giant component of
  three nodes, as depicted in Fig. \ref{model}d.  The process now
  continues back and forth between the simple percolation departure
  followed by the influence-induced departure until the cascading
  stops. For instance, one extra node departs in Fig. \ref{model}e due
  to influence leading to the final giant component of
  Fig. \ref{model}f where all the remaining influence links points
  towards nodes in the giant component and, therefore, no more
  cascading processes are possible.

The full cascading process is mathematically modeled on the equivalent
network as follows (see SI
for a detailed derivation): {\it (i)} We first apply a classical
percolation process of random removal of a node (Fig. \ref{model}a),
and remove all the nodes that become disconnected from the giant
component due to the lost of the corresponding connectivity links to
the network (Fig. \ref{model}b). In terms of generating functions,
this process leads to the set of recursive equations SI
Eqs. (\ref{s1})-(\ref{s3}) as shown in \cite{sergey}. {\it (ii)} The
next step corresponds to the process of node removal through
correlated influence links. When a node leaves the network, the
followers connected to the node via $k_{\rm in}$ links leave too
(Fig. \ref{model}c), triggering a cascading effect described by SI
Eqs. (\ref{s4})-(\ref{s6}). Notice that here we are applying this
  percolation step to the equivalent network and not the
  original. Thus, the nodes in the original network will still leave
  the network with probability $q_h<1$ when one of his $k_{\rm out}$
  followees depart.  That is, the model is mathematically solved in
  the equivalent network (where nodes leave with probability one
  following influential nodes but with smaller number of links) but
  the real dynamics is still applied to the original network with the
  original number of out-links where nodes leave the network with a
  probability $q_h<1$. {\it (iii)} This influence-induced departure
of followers can be mapped back to a second percolation removal of
nodes (Fig. \ref{model}d) and the subsequent removal by influence
(Fig. \ref{model}e). The whole process is captured by the set of
iterative SI Eqs. (\ref{mainsys5}) which describe a cascading process
that terminates when all the influence links of the nodes in the giant
component point to unremoved nodes in the same component
(Fig. \ref{model}f).

\subsection{Solving the cascading process}

 {\it (i)} The first stage in the cascading process is described
  by $\tilde{p}=p$, where $\tilde{p}$ is the fraction of links
  remaining in the giant component at a given stage in the cascading
  process. We obtain the size of the giant component $x$ and the
  survival probability $t$ of remaining nodes after the first removal
  of $p$ nodes as:
\begin{equation} x = p (1-G(t,1,1)), \label{x0} \end{equation}
 \begin{equation}
t=1-\tilde{p}(1-f), \label{t0}\end{equation} where
 $f=\frac{G_{x_1}(1-\tilde{p}(1-f),1,1)}{G_{x_1}(1,1,1)}$ and
 $G_{x_1}\equiv\partial_{x_1} G$.  The physical meaning of $t$ is that
 a node with connectivity degree $k$ has a probability $1-t^k$ to
 belong to the giant component \cite {SgPRE}.

{\it (ii)} After the first undirected connectivity percolation step,
we arrive at the second-stage removing process which is caused by
influence links. In this new process, nodes are removed if and only if
they reach any node outside the giant component following influence
links. This removing process corresponds exactly to percolation on the
directed influence network, where the  correlation $P(k,
k_{\rm in}, k_{\rm out})$ needs to be explicitly taken into
account. This process can be described by the following equations:

\begin{equation}
y=H(hx),
\end{equation}

\begin{equation}
 H(x)=x\frac{G_{x_2}(1,1,H(x))}{\langle k_{\rm in} \rangle},
 \end{equation}
where $h=\frac{\langle k_{\rm in}^x \rangle}{\langle k_{\rm in}
  \rangle}$ is proportional to the sum of the in-degree of all nodes
in $x$. These equations are derived in SI Eqs. (\ref{h1})-(\ref{s6}).
The physical meaning of $y$ is that a node with out-degree $k_{\rm
  out}$ will survive this removing process with probability $y^{k_{\rm
    out}}$. It implies that integrating the initial removing in {\it
  (i)} and these two processes in {\it (ii)} is equivalent to randomly
removing each node from the original network with probability
$1-py^{k_{out}}$.  

{\it (iii)} Therefore, the whole process {\it (i)} and {\it (ii)} can
be thought of as a single removal in the original network with the
definitions: $\tilde{p}=\frac{pG_{x_1}(1,1,y)}{G_{x_1}(1,1,1)}$ and
$f=\frac{G_{x_1}[1- \tilde{p}(1-f),1,y]}{G_{x_1}(1,1,y)}$, while $x$
and $t$ remain the same as in Eqs. (\ref{x0}) and (\ref{t0}) (detailed
derivations in SI Eqs. (\ref{pt1})-(\ref{mainsys})).

This kind of new ``initial'' removing can be described exactly by
generating function. Thus, we arrive again to stage {\it (i)} to
perform a modified undirected connectivity percolation step. The
process continues until the cascading avalanche is over.

The above analysis leads to a set of recurring relations defining
  the cascading process. After the second stage, the current cascading
  effect can be mapped to a removing process in the original
  network. This property allows us to write down the cascading process
  as recursive equations, see also SI Eqs. (\ref{mainsys5}),
which allow us to solve the whole cascading process by finding its
fixed point. 

Integrating the above three stages, we can rewrite the cascading
process as following:
\begin{align}\label{mainsys5_t}
x=p[1-G(1-\tilde{p}(1-f),1,1)],
\\\nonumber
f=\frac{G{x_1}(1-\tilde{p}(1-f),1,y)}{G_{x_1}(1,1,y)},
\\\nonumber
\tilde{p}=\frac{pG_{x_1}(1,1,y)}{G_{x_1}(1,1,1)},
\\\nonumber
t=1-\tilde{p}(1-f),
\\\nonumber
\langle k_{\rm in}^x \rangle=\frac{G_{x_2}(1,1,1)-G_{x_2}(t,1,1)}{1-G(t,1,1)},
\\\nonumber
h=\frac{\langle k_{\rm in}^x \rangle}{\langle k_{\rm in} \rangle},
\\\nonumber
y=hx\frac{G_{x_2}(1,1,y)}{\langle k_{\rm in} \rangle}.
\end{align}

The physical meaning of these recursion relations is that, after each
first stage, a node that is not removed in the initial attack has
$1-t^k$ survival probability. After the second stage, the survival
node can be mapped out to a removing process occurring on the original
network with probability $1-py^{k_{\rm out}}$. These two properties
allow us to write down the formula of the relative size of giant
component $P_{\infty}$ at the final stable state of the cascading
process at equilibrium:
\begin{equation}
P_{\infty}(p)=p[G(1,1,y)-G(t,1,y)].
\label{pinfty0}
\end{equation}

\subsection{Phase diagram}

The model predicts the existence of first-order \cite{explosive} and
second-order phase transitions. When the transition is second order,
we obtain an explicit formula for the percolation threshold (see SI
Eq. (\ref{9}) for derivation):
\begin{equation}\label{pc2}
p_c^{\rm II}=\frac{\langle k \rangle}{\partial_{x_2}^2G(1,1,0)}.
\end{equation}
Equation (\ref{pc2}) generalizes the classical uncorrelated
percolation result \cite{cohena} $p_c=\frac{\langle k \rangle}{\langle
  k (k-1) \rangle}$ to networks with influence links, and generic
correlations, $P(k, k_{\rm in}, k_{\rm out})$. The threshold for a
first-order transition, $p_c^{\rm I}$, is obtained through the
implicit formula SI Eq. (\ref{10}):
\begin{equation}
\frac{\partial t(y,p_c^{\rm I})}{\partial y}\frac{\partial y(t,p_c^{\rm I})}{\partial t}=1,
\label{tangentialattachment}
\end{equation}
where $t(y,p)$ and $y(t,p)$ are the functions describing the
influence-induced percolation process according to SI
Eqs. (\ref{10})-(\ref{12}).  The boundary between the first and second
order transitions in phase space is obtained by setting, $p_c^{\rm
  I}=p_c^{\rm II}$ leading to SI Eq. (\ref{13}).


To determine the conditions for viral cascading of influence, we
consider two cases in turn: uncorrelated and correlated networks. For
uncorrelated networks, $P_{\rm unc}(k,k_{\rm in},k_{\rm out}) = P_1(k)
P_2(k_{\rm in}) P_3(k_{\rm out})$, where the three functions are
generic probability distributions. In this case the transition is of
second order and $p_c^{\rm II, unc}$ is obtained explicitly (see SI
Eq. (\ref{highlights5})):
\begin{equation}\label{pc_unco} p_c^{\rm II,
    unc}=\frac{\langle k \rangle}{q_0
    \langle k(k-1) \rangle},
\end{equation}
where $q_0$ is the fraction of nodes with $k_{\rm out}=0$.


Surprisingly, we still find $p_c=0$ for scale-free networks (due to
the diverging second moment in Eq. (\ref{pc_unco}) for $\gamma<3$)
despite the existence of influence links.
This means that, without correlations, the influence links alone
cannot sustain a viral spreading process to break down the strong
resilience of scale-free networks; i.e., viral cascades cannot be
sustained in an uncorrelated scale-free influence network.
Indeed, empirically, we find that there exists strong 
correlations between $k_{\rm in}$, $k_{\rm out}$ and $k$: the most
active authors with large collaborative projects tend to receive and
provide the largest influence from and to their peers
(Fig. \ref{model}g).  We find:
\begin{equation}
k_{\rm out}\propto k^{\alpha}, \,\,\,\, \,\,\,\, k_{\rm in}\propto
k^{\beta},
\label{kout}
\end{equation}
 where the correlation exponents are close to one ($
 \alpha=0.91\pm0.04$ and $\beta=1.04\pm0.05$ for the APS data).
When these correlations are included in Eq. (\ref{pc2}),
we predict a non-zero correlated percolation threshold (see SI
Eq. (\ref{14})):
\begin{equation}
p_c^{\rm II, cor}=\frac{\langle k \rangle}{\sum_kk(k-1) P(k) \,
  \exp\Big({\frac{-k^\alpha \langle k_{\rm out}\rangle}{\langle
      k^\alpha \rangle}}\Big)}.
\label{pc_corr}
\end{equation}

Equation (\ref{pc_corr}) is remarkable in two aspects: First, the
value of $p_c$ increases sharply from zero as $\alpha$ increases until
a maximum value that depends on $\gamma$ and $\langle k_{\rm out}
\rangle$ (Fig. \ref{diagram}a).
The vulnerability increases until the term $\langle k^\alpha \rangle$
becomes dominant and stabilizes the network with the concomitant
decrease of $p_c^{\rm II, cor}$ back to zero as $\alpha\to\infty$.
Second, $p_c^{\rm II, cor}$ is independent of $\beta$, since $x_2=1$
in Eq. (\ref{pc2}), implying, rather surprisingly, that the influence
exerted by the large number of ingoing influence links of the hubs is
not enough to produce viral spreading.

The theoretical results are tested against computer simulations
  of numerically generated scale-free networks with a prescribed set
  $(\alpha, \beta, \gamma, \langle k_{\rm out} \rangle)$. We first
  generate a scale-free network with a given value of $\gamma=2.5$ and
  minimal degree equal to one.  For a give node with connectivity
  degree $k$, the influence out-degree is proportional to $k^\alpha$
  and the in-degree is proportional to $k^\beta$. We choose the number
  of in and out influence link from a Poisson distribution with
  average given by these two values. The Poisson distributions
  $P(k_{\rm out}|k)$ and $P(k_{\rm in}|k)$ are validated for the five
  APS communities in SI-Fig. \ref{poisson}.  Thus, we generate the
  connectivity network and the correlated influence directed links
  according to the connectivity degree of each node and $(\alpha,
  \beta)$. We then calculate numerically $P_\infty(p)$ by performing a
  percolation cascading process directly on the network. Then, we
  calculate $G(x_1,x_2,x_3)$ to obtain the theoretical predictions of
  Eqs. (\ref{pinfty}) and (\ref{9}). Figure \ref{model}h shows the
  comparison between the theoretical results and the simulations.  We
  obtain very good agreement between the predicted $P_\infty(p)$ and
  $p_c^{\rm II}$ and the numerical estimation obtained by applying a
  percolation process to a correlated network with influence links
  with the parameters expressed in the figure.  

 It is important to note that the generating function formalism is
 based on a locally tree-like assumption.  Such an assumption is
 satisfied locally in random networks as well as scale-free networks
 and this is the reason of the very good agreement between theory and
 simulation in Fig. \ref{model}h. However, real networks are not
 tree-like and local clustering is an important property of any real
 world network. For instance, clustering in complex networks can
   be classified in two different classes, weak and strong. Strong
   clustering occurs where triangles in the network share edges, so
   that the multiplicity of edges can be high. Weak clustering occurs
   when triangles do not share edges.  A formalism for weakly
   clustered networks has been recently considered in
   \cite{newman_clustering}.  However, strong clustering occurs more
   often in real networks.  Thus, a more realistic theory would need
   to considered to capture the existence of strong clustering in
   real-world systems.

Taken together, these results paint a picture of a viral cascading
process where few small players--- not the hubs--- initiate the
cascades. However, the hubs play a key role in sustaining the
cascades, not as pioneers but as followers: since the well-connected
nodes receive a greater influence (via $k_{\rm out} \sim k^\alpha$),
they are more ``aware'' of the latest developments. This allows the
hubs to ``jump'' to the new trend easily. In percolation terminology,
the random removal of nodes (which targets mostly low-degree
individuals) becomes, at a later stage in the cascade, a targeted attack
on the hubs via their large number of outgoing influence links. This
is due to the cascading effect where low-degree pioneering nodes can
now have easy access to the well-connected hubs through their large
number of $k_{\rm out}$ links.  This effect explains the condition for
the catastrophic fragility and the viral spreading in the highly
correlated influence network. Contrary to expectations, the large
ingoing influence of the hubs (via $k_{\rm in} \sim k^\beta$) plays no
role in sustaining the cascade.

\subsection{Testing the theoretical results}

We test our theoretical predictions by calculating $P_\infty(p)$ from
Eq. (\ref{pinfty0}), and comparing with the collaborative networks of
Fig. \ref{important_figure}a (a comparison with LiveJournal is
performed in the next section).   The statistical estimation
  procedure of parameters (via, for instance, standard maximum
  likelihood of power-law distributions) to produce inputs to the
  mathematical model is a drawback of the modelization. Indeed, the
  mathematical model has probabilistic underpinnings, and therefore
  estimation based directly on those underpinnings would be more
  appropriate. Therefore, we directly use the empirical degree
  distribution into the theory to provide the theoretical estimation
  of the giant component in the $(p,P_\infty)$-plane in
  Fig. \ref{important_figure}a. 


Additionally, current approaches in the statistical literature
\cite{ERGMWasserman1} model the observed links of networks as the
fundamental level of data hierarchy through which network structure is
considered. Thus, rather that using an amalgamated data, such as
degree distribution, for the base level of the model, we perform
statistical analysis in terms of the exponential random graph models
(ERGM) \cite{ERGMWasserman1} using
stochastic algorithms based on Markov chain Monte Carlo (MCMC) method
to allow for the use of statistics (degree distribution) of a network
as predictors of links. This method further allows for the estimation
of distributions of functions based on a model for a network
(predictive distributions).

For the exponential-family random graph model, we define
$P_\theta(\Omega=\omega) = \exp[\sum_{k=0}^{n}\theta_k p_k(\omega)]$,
where $p_k(\omega)$ is the probability of randomly choosing a node
with degree $k$ in a network $\omega$. First, we use the empirical
network to estimate all the parameters $\theta_k$.  Then, we use MCMC
to generate 100 networks with the same features captured by the model.
We use the professional software {\it ERGM} obtained at
http://statnetproject.org to estimate the parameters $\theta_k$ and
generate the 100 networks \cite{Hunter} (more details in SI).

We then employ a bootstrap method to estimate the exponent $\gamma$ of
the degree-distribution.  Employing bootstrap method combined with
maximum-likelihood estimation and Kolmogorov-Smirnov test
\cite{maxlikehood}, we obtain the exponent of the degree distribution
as $\gamma = $2.97 (0.01 standard deviation).  

Furthermore, we use $(\alpha, \beta) = (0.91, 1.04)$
obtained from the real data.
We also estimate the lower and upper bounds of the average influence
degree $\langle k_{\rm in} \rangle=\langle k_{\rm out} \rangle$ from
the data (see SI).
The empirical lower bound is $\langle k^{\rm equiv}_{\rm out} \rangle
= 0.44$, which gives rise to the predicted $P_\infty(p)$ shown as the
red curve in Fig. \ref{important_figure}a with percolation threshold
0.54.  The figure shows that this theoretical prediction provides a
lower bound to the empirical data, providing support for the
model. For larger $\langle k_{\rm out} \rangle $--- for fix ($\alpha,
\beta$)--- the vulnerability of the network increases according to
larger percolation thresholds (inset
Fig. \ref{diagram}a). Furthermore, a second-order transition at small
$\langle k_{\rm out} \rangle$ turns into an abrupt first-order
transition at large $\langle k_{\rm out} \rangle$, as shown in the
phase diagram of Fig. \ref{diagram}b.  As $\langle k_{\rm out}
\rangle$ increases, the threshold condition changes from
Eq. (\ref{pc2}) to Eq. (\ref{tangentialattachment}), and the
transition becomes discontinuous. For instance, while for $\langle
k_{\rm out} \rangle = 0.44$, the transition is just at the boundary
between the first order and second order transition (red dot in
Fig. \ref{diagram}b), for $\langle k_{\rm out} \rangle = 0.83$, the
transition becomes first-order.  The values $\langle k_{\rm out}
\rangle = 0.44$ and 0.83 provide a lower and upper bound of the
empirical data $P_\infty(p)$ as shown in the red and blue curves in
Fig. \ref{important_figure}a, providing further support for the
model. The first-order scenario implies a dramatic viral spreading
where a network near its threshold will suddenly disintegrate by the
departure of an infinitesimal number of its members. Since the
empirical data in Fig. \ref{important_figure}a is close to the upper
bound provided by $\langle k_{\rm out} \rangle = 0.83$ (specially the
fields Phase Transitions, Fluctuations and Interfaces), it is
plausible that these scientific communities are being disintegrated by
catastrophic discontinuous events.

We would like to note that we do not claim to fit the five
  datapoints on the APS communities with a single functional form
  obtained from the theory. In fact, each point may corresponds to an
  independent percolation process given by a different $\langle k_{\rm
    out} \rangle$. Instead, we show that these five datapoints are
  within the upper and lower bounds of the theory. Indeed, later we
  will repeat the same analysis to the LJ communities. These
  communities are much larger in number totaling 10,981 LJ
  communities. The data for this large number of communities is
  strikingly well fitted by the theory as we will see in
  Fig. \ref{lj}.




One assumption of the model is that the scientists who leave a
  declining network, are leaving due to the influence exerted by the
  pioneers.  However, scientists might be leaving a field for a
  variety of reasons such as a declining field and also to other
  destinations.

To study these questions, we have measured the ratio between departing
scientists who moved to Complex Networks to the total number of
departing scientists from a given field. The percentages for the
different fields are Fluctuation: $86.0\%$, Chaos: $88.1\%$,
Thermodynamics: $95.7\%$, Phase Transitions: $85.1\%$ and Interfaces:
$36.4\%$. Thus, except for the field of Interfaces, the other fields
show a large percentage of departing scientists going to Complex
Networks.

Yet, the fact that scientists leave a field to work in Complex
Networks does not necessarily mean that they are following the
pioneers. That is, the motives behind such a decision could be
different and attributing the full attrition of a field to the
influence of pioneers implies an assumption of causality which may not
be satisfied. For instance, a scientist may leave the field under the
impression that the original field is declining.

While the definite answer to this question would require a survey to
know actual motives of scientists, the good agreement between model
and empirical data (including the LiveJournal network studied next)
suggests that the influence percolation model reproduces well the
disintegration of communities. This result, in turn, suggests that an
important mechanism behind the network disintegration are the
correlated influence links. 

 We note that, while the model assumes local correlations
 between the different degrees of a given node, the correlation
 between the connectivity and influence networks themselves are
 neglected.  To test for these correlations, we measure
 the average influence degree (in and out) of the nearest neighbors,
 $\langle k_{\rm nn} \rangle$, of a node with connectivity degree $k$
 connected via an influence link in the APS networks (SI
 Fig. \ref{knn}). We find a lack of correlations indicated by a flat
 $\langle k_{\rm nn} \rangle$.  If such a correlation exists, it may
 contribute to an under-estimation of the cascade effect. Therefore in
 SI we develop an extension of the theory to treat these correlations
 as well.

\subsection{Disintegration of LJ communities}

 The mathematical modeling assumes that the settings are mutually
  exclusive, while in practice this may not be true in the example of
  the APS communities. For this reason, we also tested the model on
  another dataset: the communities formed by bloggers in
  LiveJournal.com (all datasets are available at
  http://lev.ccny.cuny.edu/$\sim$hmakse/soft\_data.html).  

LiveJournal (LJ) is a large social network of 8.3 million users who
post information and articles of common interest. This community has
been used in network studies of information flow and influence
\cite{lev,livejournal}, since it was shown to have features consistent
with other large-scale social networks.

 We have recorded the posts in the LiveJournal social network from
 February 14th, 2010 to November 21st, 2011.  We have also sampled the
 full network of LJ users and also the declared interest of each user
 which defines the community to which the user belongs. Data
 collections on the network has been performed every 1.5 months so
 that we have 14 snapshots of the entire LJ structure. The entire
 history of posts of users have been recorded continuously over the
 studied period of time. This information allows us to define the
 variables that are used in the model to describe the disintegration
 of communities:  (a) The connectivity network: when users $i$
 and $j$ declare their friendship in the network,  (b) The
 influence network: when user $i$ cites posts of user $j$, then we
 consider a directed influence link from $i \to j$ since user $i$ is a
 follower of user $j$. Thus, we can define the respective degrees of
 connectivity links $k$ and influence links, $k_{\rm in}$ and $k_{\rm
   out}$ and search for correlations between these
 variables.   (c) Finally, each user in LJ declares a community
 to which the user belongs (sports, literature, etc).

Therefore, we have the three main ingredients of the theory:
connectivity and influence links and well defined communities that we
can track over an extended period of time.  Crucially, we are able to
track those communities that are created and disintegrated in the
number of 10,981.  In LJ, the communities are declared by the users,
and users change interest very often, creating and disintegrating
communities quite often. In this case, the settings may be mutually
exclusive, since the communities appear to be very dynamic and users
change interests rapidly.

The analysis of the LJ communities reveals a remarkable result: In
Fig. \ref{lj}a we study the size of the giant components $P_\infty$ of
the disintegrated networks to produce the percolation plot of
$P_\infty$ as a function of the leaving pioneers $1-p$. We find that
the disintegration of the LJ communities follows closely a quite
generic curve in the $(p,P_\infty)$-plane indicating rapid
disintegration and great fragility of the communities via cascading
effects with critical percolation threshold $p_c = 0.962$. The
empirical curve can be well-fitted by the theoretical model
(Fig. \ref{lj}a) when similar local correlations between
connectivity and influence links as in the APS communities are taken
into account (Fig. \ref{lj}b).  Our results are consistent with an
abrupt first-order transition occurring during the disintegration
process as indicated in the phase diagram of Fig. \ref{diagram}b.


\section{Discussion}

From the development of new ideas, brand-new products to political
trends, the present model shows the conditions for viral spreading:
when $k_{\rm out}$ and $k$ become highly correlated (large $\alpha$),
a few individuals, who are not necessarily the hubs, can trigger a
large cascade leading to network fragmentation.  In conclusion,
through mathematical and empirical calculations, we establish two
emergent properties that result when overlying multiplex
networks interact:

(i) We mathematically derive the necessary conditions for sustaining a
viral spreading process. We show how damage in a network can, in turn,
damage the influence network and vice versa, leading to viral cascades
of followers. Our modeling predicts the conditions for these
interactions to sustain a large viral spreading via a precise scaling
form of the correlation function between multi-layers,
Eqs. (\ref{kout}).

\begin{figure*}[hbt]
\centering {\includegraphics[width=.9\columnwidth]{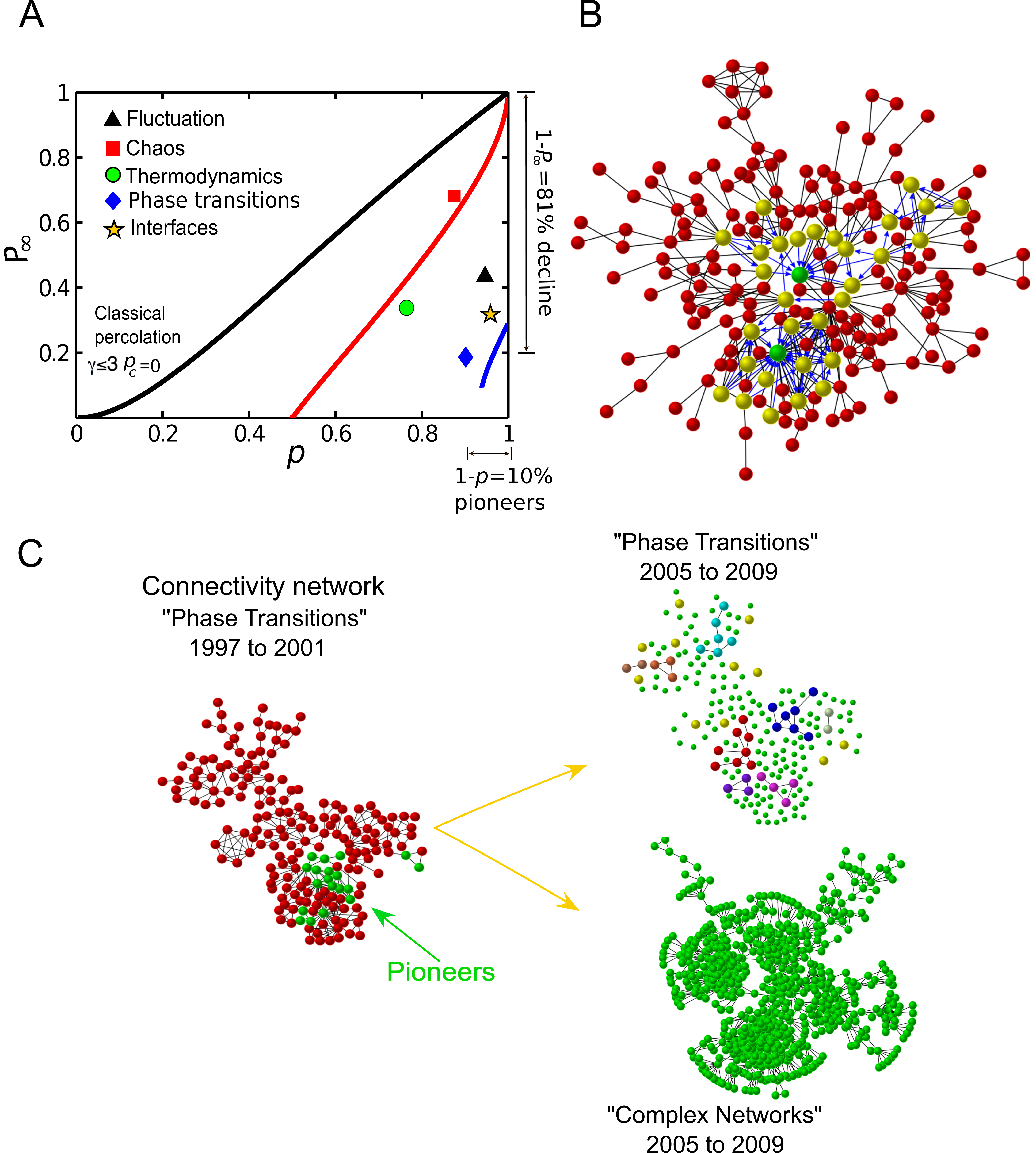}}
\centering 
\includegraphics[width=0.5\columnwidth]{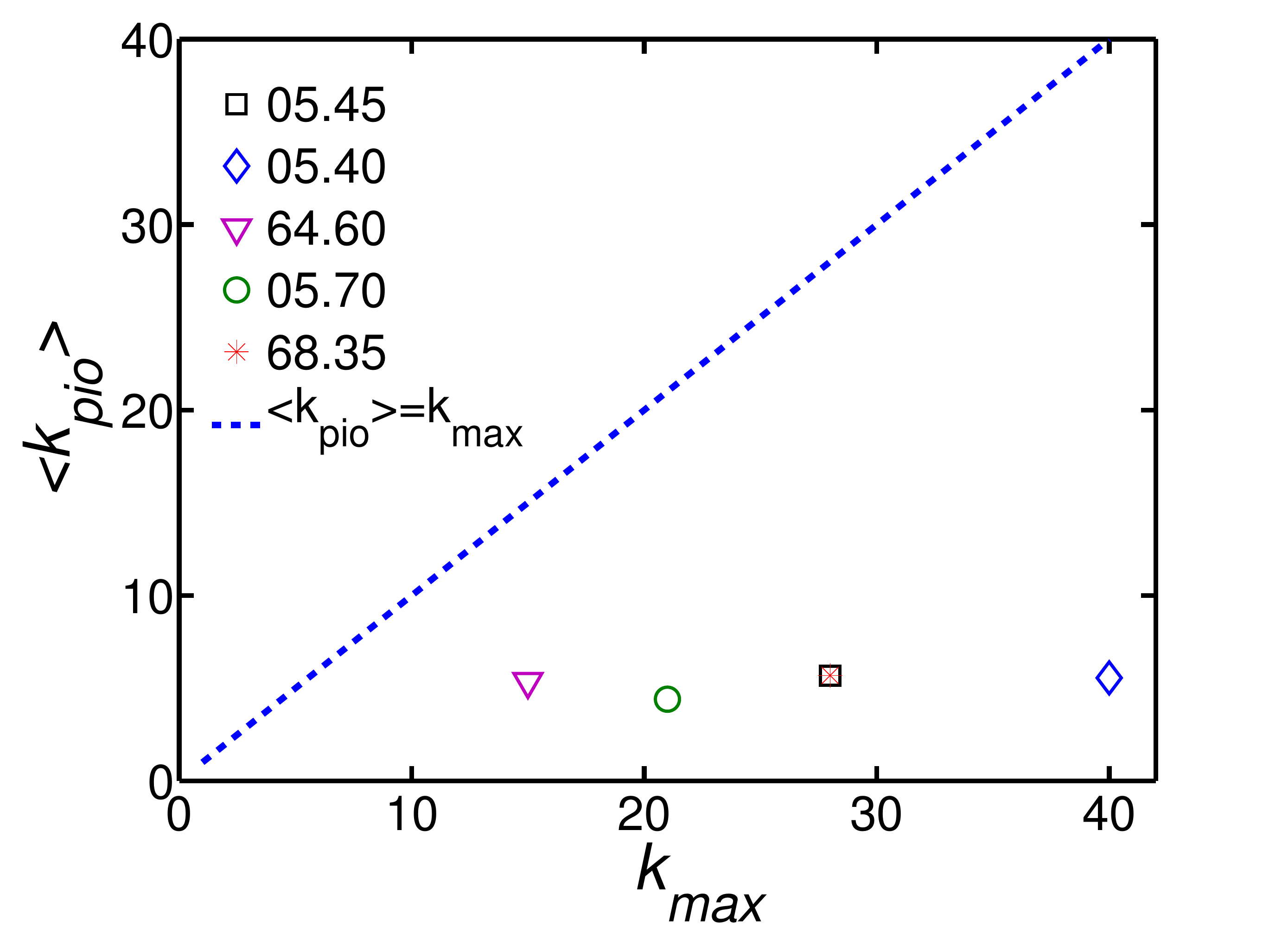}
(D)
\caption{}
\label{important_figure}
\end{figure*}

\begin{figure*}[hbt]
\centering 
  \includegraphics[width=0.8\columnwidth]{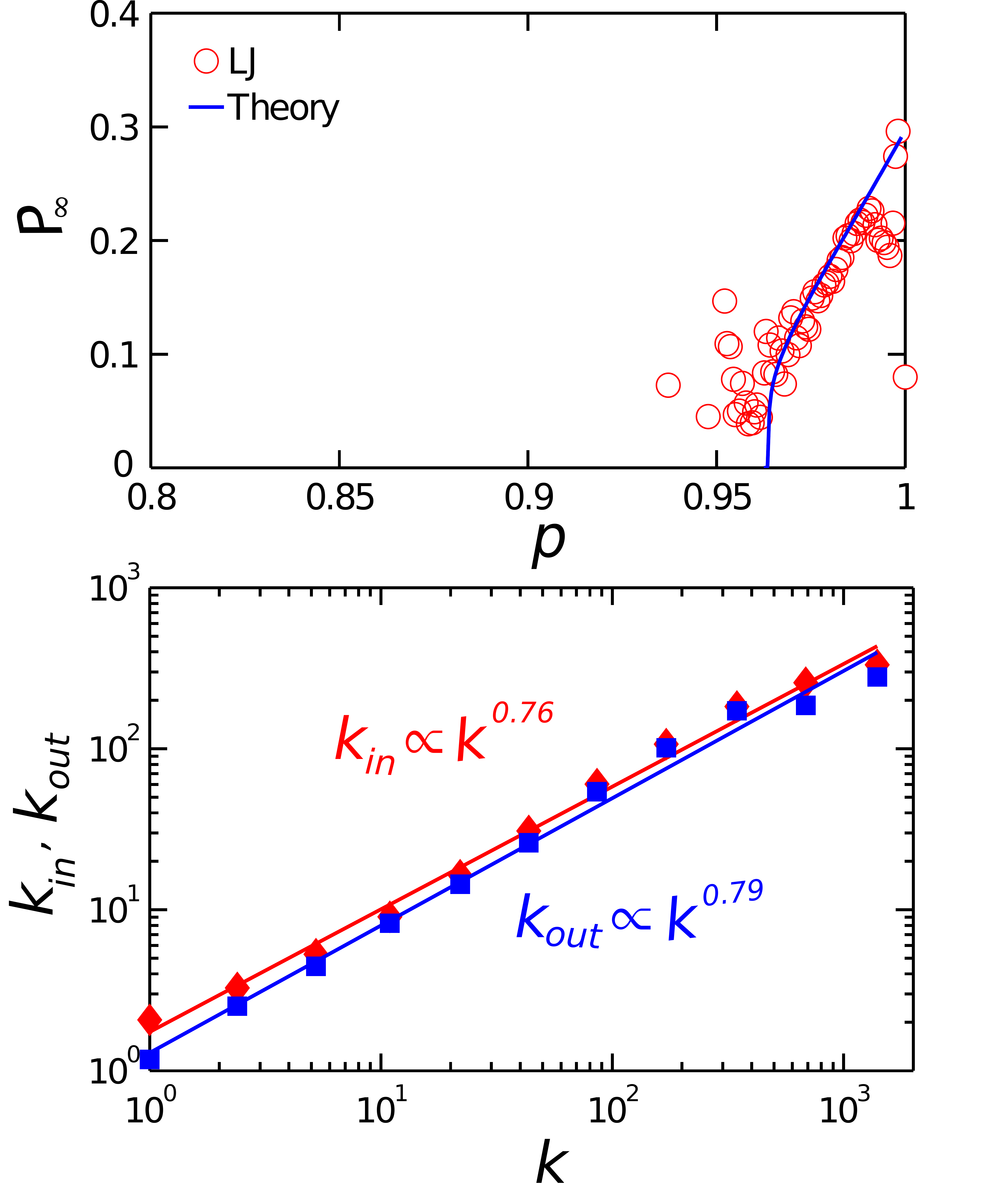}
\caption{}
\label{lj}
\end{figure*}

(ii) This theoretical prediction is in agreement with our empirical
observations (Fig. \ref{important_figure}a and \ref{lj}a). We find
that the conditions for viral spreading, Eqs. (\ref{kout}), to be
valid in the studied networks.  This viral effect is empirically
quantified with the large percolation threshold ($p_c$ in
Fig. \ref{important_figure}a and \ref{lj}a) which is also predicted by
the theory.  Contrary to expectation, the innovators are not the hubs,
but the small players.

A related question arises whether there is a universal model to
  explain the fall of all kinds of communities/trends/topics even when
  characterized by different time scales. Such a model would be
  difficult to implement. Here we show that in two different networks,
  disintegration of communities can be understood via a modified
  percolation model in a multiplex network including correlated
  influenced links. These two networks have different time scales for
  disintegration where communities rise and fall in a matter of weeks
  (LJ) to years or even decades as in scientific trends in
  science. Certainly we have not exhausted all the cases, but the
  present data is indicative enough to suggest that the same modeling
  could be applied to other networks. In particular, it could be
  applied to Facebook of Tweeter where trends appear and disappear in
  similar fashion as in LJ.

Our results have consequences for a range of social, natural and also
engineered systems. They cause us to rethink the assumptions about
robustness and resilience of social networks, with implications for
understanding viral spreading in social systems and the design of
robust multiplex interconnected networks.  In the present study,
we have tested the theoretical results on typical cases of scientific
collaboration network and online information dissemination, but the
results are equally applied to a variety of interconnected 
  multiplex systems with correlated influence links. These systems
range from political networks, to financial markets and the economy at
large.

Acknowledgements. This work was supported by ARL
under Cooperative Agreement Number W911NF-09-2-0053, NSF-PoLS and NIH.
SH wishes to thank LINC and Multiplex EU projects and the Israel
Science Foundation for support.  We thank S. Alarc\'on, A. Bashan,
L. K. Gallos and S. Pei for useful discussions, M. Doyle and
H. D. Rozenfeld for providing the APS dataset and L. Muchnik for
providing the LiveJournal dataset.

\pagebreak

\begin{figure*}[hbt]
\centering{ \includegraphics[width=0.9\columnwidth]{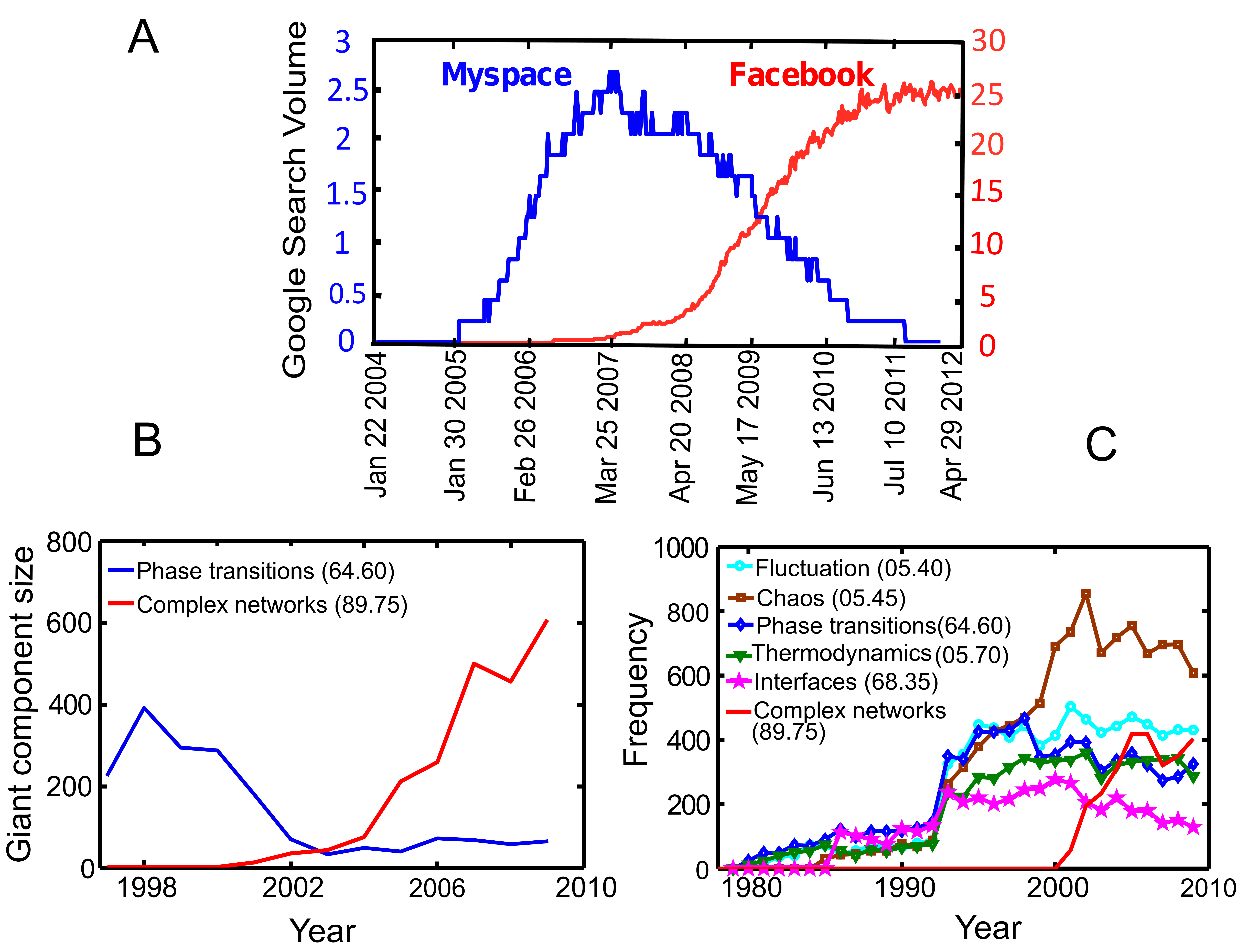}}
\caption{}
\label{PercolationNetwork}
\end{figure*}

FIG. \ref{PercolationNetwork}.  Rise and fall of
  communities. (a) Comparison of activity in ``Myspace.com'' vs
``Facebook.com'' from 2004 to 2012 through Google's Search Volume
Index \cite{trends}, which measures the number of Google searches of
each word.  The fall of myspace coincided with the rise of Facebook
suggesting that users moved from one network to the other.  The
tipping point can be identified on March 25, 2007. (b) The size
of the largest (giant) connected components of scientists studying
``Phase Transitions'' (PACS 64.60), and ``Complex Networks'' (PACS
89.75) from 1997 to 2009.  The steady increase in the study of Phase
Transitions declined as the Complex Networks field started to grow in
the physics community.
(c) The frequency of citations per year of the top five fields
contributing scientists to the rise of Complex Networks.

FIG. \ref{important_figure}.  Cascades of followers through the
  influence of pioneers in APS.   (a) Relative size of the
largest component, $P_\infty(p)$, of collaborating scientists in the
indicated fields after the departure of $1-p$ pioneers to the field of
Complex Networks ($s=2$, other values yield similar results).  The
solid curves denote the different theoretical predictions. Black curve
(extreme resilience): classical percolation theory on a scale-free
network predicting $p_c=0$
\cite{cohena,satorras01,barabasi-attack,NewmanSpread}. Red curve (high
vulnerability): prediction of influence-induced correlated percolation
for $(\alpha, \beta, \langle k_{\rm out} \rangle) = (0.91, 1.04,
0.44)$
with a predicted threshold very close to the boundary between first
and second order transition $p_c^{\rm I} \approx p_c^{\rm II} =
0.54$. Blue curve (extreme vulnerability): prediction of
influence-induced correlated percolation for $(\alpha, \beta, \langle
k_{\rm out} \rangle) = (0.91, 1.04, 0.83)$ giving rise to a
first-order transition with $p_c^{\rm I}=0.97$.  This means, the
departure of $3\%$ of pioneers will cause cascading followers and
collapse of the original network.  The red and blue curves are bounds
to the real data.  (b) The influence network (blue links) of
collaborating scientists in the field of Phase Transitions.  Green
nodes are a sample of pioneers of the field of Complex Networks and
the yellow nodes are their closest followers that departed
afterwards. It is apparent the large cascading effect produced by the
departing nodes. Black links are connectivity links.   (c) The
largest connected component of the collaboration networks of
Phase Transitions up to 2001 and its reduced state in 2005-2009 with
the concomitant creation of Complex Networks.  (d) Pioneers are
not the hubs.  The average degree of pioneers in the largest
cluster, $\langle k_{\rm pio} \rangle$ versus the maximum
connectivity degree $k_{\rm max}$ over the members of each PACS
field. We find that in general the degree of the pioneers is much
smaller than the maximum degree indicating that the pioneers are not
the hubs (see also SI Table \ref{table1}).

\begin{figure*}[hbt]
\centering 
\includegraphics[width=0.6\columnwidth]{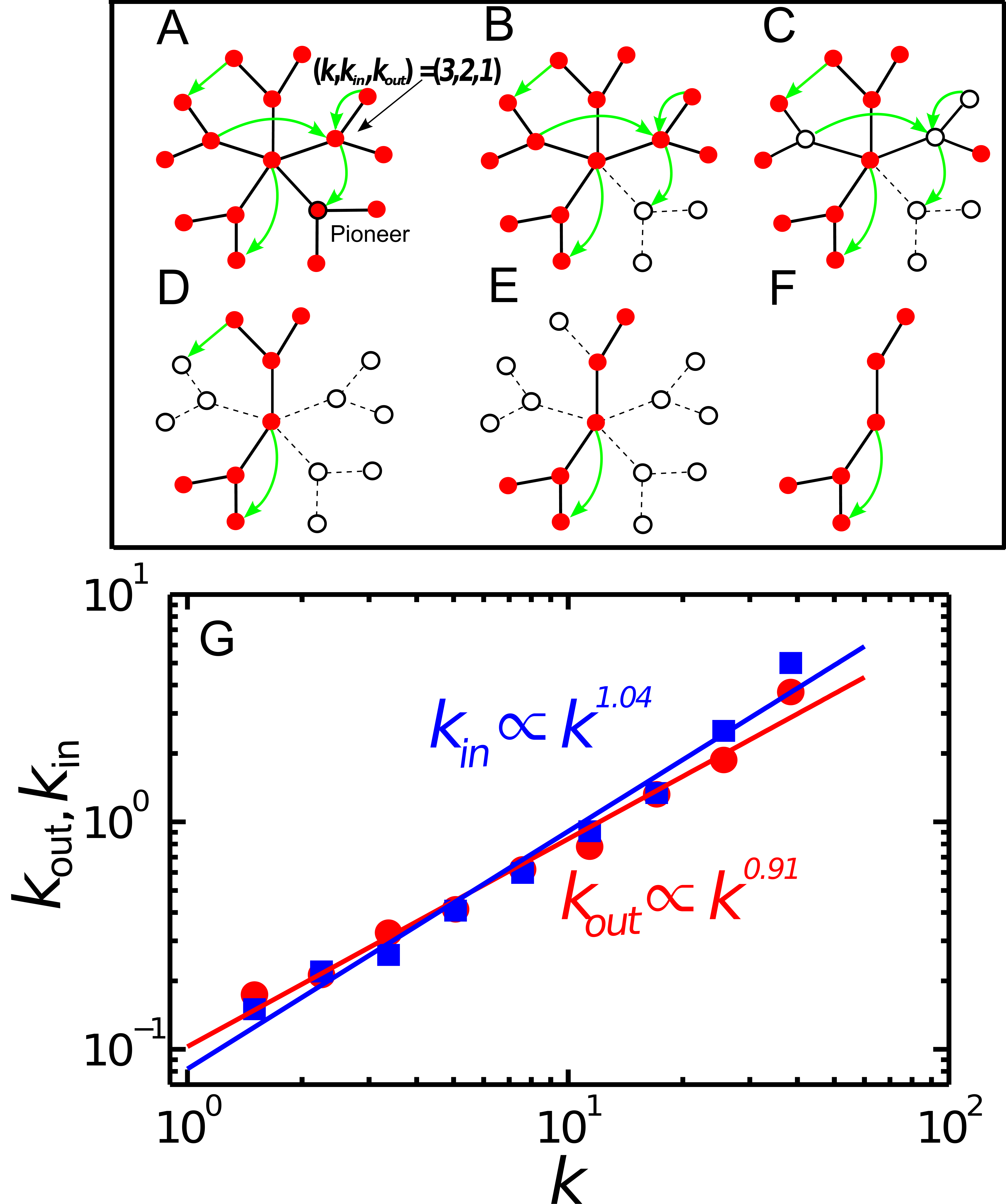}
\centering
\includegraphics[width=0.6\columnwidth]{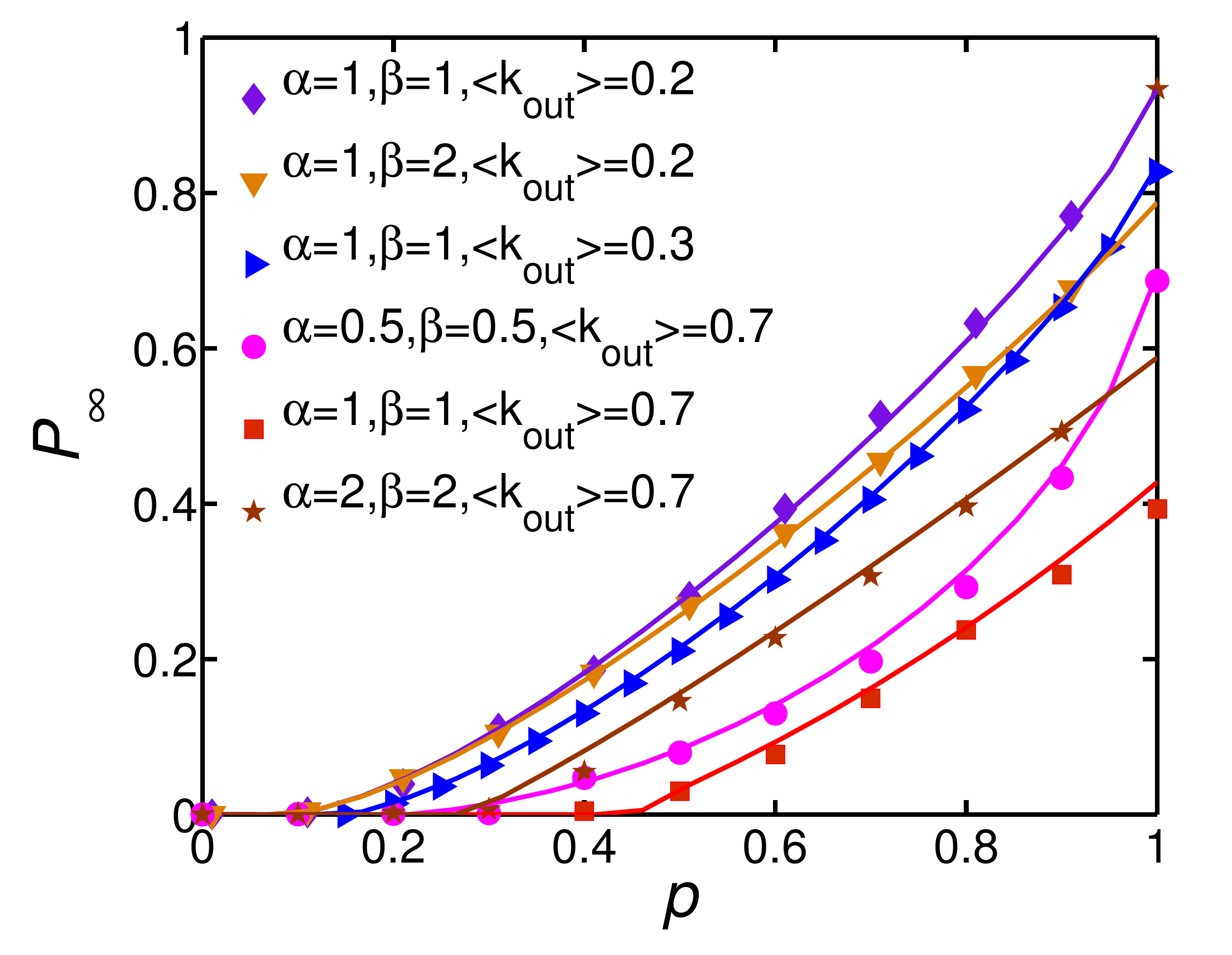}(H)
\caption{}
\label{model}
\end{figure*}

FIG. \ref{model}.  Modeling the cascading process of
  followers.  (a) Sketch of the model network (considered
  as the giant component) with undirected connectivity links (solid
  black links) superimposed by a network of directed influence links
  (green links). The nodes are characterized by ($k, k_{\rm in},
  k_{\rm out})$ as indicated in the sample node.  The node indicated
  by the black circle is the pioneer moving to another field and is
  therefore removed first.  (b) First connectivity percolation
  process: two additional nodes are removed (open circle) since they
  are not connected any more to the giant component after the removal
  of the pioneer node in  (a).  (c) Influence-induced
  process: one extra node is removed due to the influence link
  pointing to the pioneer which induces two more influence departures
  as indicated. Such removals induce a cascading effect since now
  other nodes are disconnected from the giant component; a step that
  is considered next.  (d) Percolation connectivity process:
  three extra nodes are disconnected from the giant component and are
  therefore removed.   (e) Influence-induced percolation process:
  a final node is removed due to the influence link to one of the
  nodes removed in (d).  (f) At the end of the cascade,
  the giant component is reduced to six nodes.   (g) Empirical
study of local correlations between influence degree and
connectivity degree averaged over all APS networks. We obtain: $k_{\rm
  out}\propto k^{\alpha}$ and $k_{\rm in}\propto k^{\beta}$ with
$\alpha=0.91\pm$0.04 and $\beta=1.04\pm$0.05. (h) Comparison
between simulations and theoretical results. The symbols denote the
simulation results and the curves are the prediction of theory for
$P_\infty(p)$.  We use a scale-free network with $\gamma = 2.5$ and
minimal degree 1.  We first generate the connectivity network and then
generate the correlated influence directed links according to the
connectivity degree of each node and $(\alpha, \beta)$ and calculate
$P_\infty(p)$ by performing a percolation analysis directly on the
network. Then, we calculate $G(x_1,x_2,x_3)$ to obtain the theoretical
predictions of Eqs. (\ref{pinfty}) and (\ref{9}). We find that the
theoretical results agree very well with the simulations.

\begin{figure*}[hbt]
\centering 
\includegraphics[width=1.1\columnwidth]{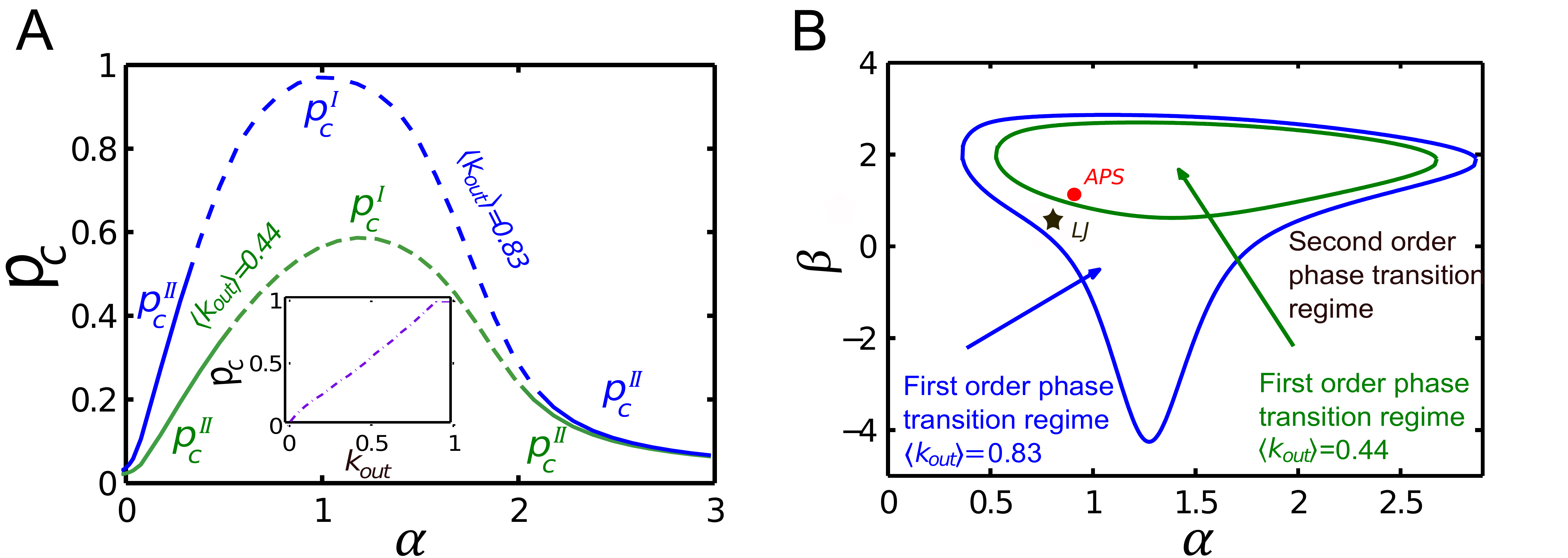}
\caption{}
\label{diagram}
\end{figure*}

FIG. \ref{diagram}.   The phase diagram predicted by the influence
  percolation model with correlation. (a) Prediction of the
percolation threshold versus $\alpha$ according to
Eqs. (\ref{tangentialattachment}) and (\ref{pc_corr}) for
$(\beta,\gamma)=(1.04, 2.90)$ and $\langle k_{\rm out} \rangle$ as
indicated. Solid lines denote the region in $\alpha$ of second order
transitions, while dashed lines denote first-order transitions. Inset
shows the increase of $p_c^{\rm II}$ with $\langle k_{\rm out}
\rangle$ for $(\alpha, \beta, \gamma)=(0.91, 1.04, 2.90)$.   (b)
Phase diagram denoting the areas in the plane $(\alpha,\beta)$ of
first-order and second-order regimes for two values of $\langle k_{\rm
  out} \rangle$ as indicated. The first order regime is inside the
indicated curve, while the second order is outside for a given value
of $\langle k_{\rm out} \rangle$. The location of the APS networks in
the phase diagram is indicate as a red dot. The networks are located
near the boundary of the transitions for $\langle k_{\rm out}\rangle =
0.44$ and inside the region of first-order for $\langle k_{\rm
  out}\rangle = 0.83$.  The LJ network is also located in the
first-order transition regime.

FIG. \ref{lj}.  Disintegration of communities of bloggers in
  LiveJournal.com. (a) Percolation plot in the plane
$(p,P_{\infty})$ of the declining communities.  The communities follow
a generic curve in the plane $(p,P_\infty)$ quantifying the rapid
disintegration and cascading effects.  We find a very good agreement
with the empirical data suggesting that LJ communities disintegrate
following a first-order transition (see also Fig. \ref{diagram}b for
the position of LJ in the phase diagram).   (b) Similarly to the
results on the APS communities, we find positive local
correlations of out and in influence degree and connectivity degree
with exponents $\alpha =0.79\pm0.03$ and $\beta=0.76\pm0.03$,
respectively.


\clearpage 

\clearpage

\centerline{ \bf SUPPLEMENTARY INFORMATION}

\vspace{0.4cm}

\centerline{Conditions for viral influence spreading through
  correlated multiplex networks}

\vspace{0.2cm}
\centerline{Yanqing Hu, Shlomo Havlin, Hern\'an A. Makse}

\vspace{.4cm}

\section{General properties of studied networks}
\label{general}


We consider the top five PACS numbers contributing scientists to the
Complex Networks field. The PACS numbers are listed in Table
\ref{table1}. The whole dataset is made available at http://jamlab.org
and it was provided by APS. The field of ``Complex Networks'' started
in 2001 in the Physics community and encompasses a series of PACS
numbers under the general class: 89.75, ``Complex
systems''. This generic PACS number contains the subclasses: 89.75.Da,
``Systems obeying scaling laws''. 89.75.Fb, ``Structures and
organization in complex systems'' 89.75.Hc, ``Networks and
genealogical trees''. The newly developed community of ``Complex
Networks'' published under the above PACS numbers.
To test this assumption we have checked all APS papers with
PACS Number 89.75 from 2000 to 2009.
We have checked that the titles of 1193 papers in PACS 89.75 contain
at least one of these words: ‘''network'', ``networks'', ``graph'',
``graphs'', ``link'', ``links'' and ``degree''. Then, it is reasonable
to assume that this PACS number has been assimilated by the newly
formed community of Complex Networks. Indeed, the present paper will
be archived under PACS 89.75.

According to the record in the APS database, we find all the
scientists who at least published one scientific paper with PACS
number 89.75 from 2001 to 2009, then go back to the period 1993 to
2000 and count the frequency of all PACS number used by these
scientists. Thus we can detect the order of each PACS number
contributing to network science. The top five PACS numbers
contributing to Complex Networks are (ordered from large to small):

\begin{enumerate}
\item  05.45. Nonlinear dynamics and chaos. Includes:
  Low-dimensional chaos, Fractals, Control of chaos, applications of
  chaos, Numerical simulations of chaotic systems, Time series
  analysis, Synchronization; coupled oscillators, etc.

\item 05.40. Fluctuation phenomena, random processes, noise, and
  Brownian motion. Includes: Noise, Random walks and Levy flights,
  Brownian motion.

\item 68.35. Solid surfaces and solid-solid interfaces. Include:
  Interface structure and roughness, Phase transitions and critical
  phenomena, Diffusion; interface formation, etc.

\item 64.60. General studies of phase transitions. Includes:
  Specific approaches applied to studies of phase transitions,
  Renormalization-group theory, Percolation, Fractal and multifractal
  systems, Cracks, sandpiles, avalanches, and earthquakes, General
  theory of phase transitions, Order-disorder transformations,
  Statistical mechanics of model systems, Dynamic critical phenomena,
  etc.

\item 05.70. Thermodynamics. Includes: Phase transitions:
  general studies, Critical point phenomena, Nonequilibrium and
  irreversible thermodynamics, Interface and surface thermodynamics.

\end{enumerate}

\begin{table}[ht]
\caption{General properties of the APS communities according to their
  PACS numbers. Values are calculated for $s=2$. $N_\infty$ is the
  size of the largest connected component calculated from 1997 to
  2001, $n_\infty$ is the size of the largest component from
  2005 to 2009, $\langle k_{\rm pio} \rangle$ is the average
  connectivity degree of the departing pioneers in 2001, $k_{\rm max}$
  is the maximum connectivity degree, $\langle k \rangle$ is the
  average connectivity degree.}  \centering
\begin{tabular}{l c c c c c c c}
\hline\hline Field & PACS & $N_\infty$ & $n_\infty$ & $\langle k \rangle$ & $\langle
k_{\rm pio} \rangle$ & $k_{\rm max}$
\\ \hline
1. Chaos&05.45&1126&846&4.26&5.56&40\\
2. Fluctuations&05.40&522&227&3.97&5.68&28\\
3. Interfaces&68.35&232&77&5.64&4.14&21\\
4. Phase transitions&64.60&193&36&3.84&5.69&28\\
5. Thermodynamics&05.70&87&32&3.70&5.36&15\\
Complex networks&89.75&9&190&-&-&-\\ [1ex]
\hline
\end{tabular}
\label{table1}
\end{table}

For each APS network identified by the above PACS numbers, we
calculate the size of the largest connected component $N_\infty$
using the coauthorship in papers published over a window of 5 years
from 1997 to 2001 using the threshold $s=2$ (the results are similar
with $s=1$ and $3$). That is, we consider scientists who published at
least $s$ papers together from 1993 to 2009.
Using the publication information of 2001 we detect the fraction $1-p$
of authors who published at least one paper in Complex Networks. We
then use a new 5 years window to define a network from 2005 to 2009
for each PACS number, and calculate the new size of the  largest
component, $n_\infty$. In these five years, a scientist may publish
papers with the given PACS number and 89.75 or both. In this case, we
classify the paper into 89.75 when the scientist published with 89.75
more times than other PACS numbers during the five year period.
Thus, we obtain the relative size of the largest component
$P_\infty(p)=n_\infty/N_\infty$ for each PACS field, which represents
a measurement of the probability that a randomly chosen node belongs
to the spanning cluster in percolation theory. The calculation is
performed for $s=2$ and is plotted in Fig. \ref{important_figure}a
(other values of $s$ yield similar results).



LiveJournal.com.-- LiveJournal (LJ) is a large social
  network of 8,312,972 users who post information and articles of
  common interest.  In LJ, each user maintains a friends list, which
  declares social links to these individuals. The network resulting
  from these social links is believed to reliably represent actual
  social relations of LJ users.  This network represents the 
    connectivity links in our model.  More importantly, since we know
  the entire history of posts over several years, we can use this
  information to quantify the influence links between two users: if
  user $i$ cited repeatedly the posts of user $j$, then we consider an
  influence link from $i \to j$ since $i$ is a follower of $j$.
  In LJ, the communities are declared by the users, and users change
  interest very often, creating and disintegrating communities quite
  often. In the two years period when we recorded LJ, we find that
  there are 10,981 communities that are being disintegrated.  We
  analyze only communities with more than 300 members, the largest of
  which contains almost half of LJ.  We have the full data on the
  network connectivity and the influence connectivity of LJ from
  February 14th, 2010 to November 21st, 2011. To calculate the
  percolation plot $(p,P_{\rm infty})$ we consider the size of the
  giant component of a declining community at the beginning of the
  period of observation and compare with the size of the same
  community at the end of the observation window.

The communities in LiveJournal evolve at a much faster pace than the
scientific communities in APS. Indeed, we need many years (the APS
dataset spans 35 years) to see an idea to be established in the
scientific community. Further, many scientists keep working in their
old fields, while moving to a new one. However, the pace of change is
much faster in an online blogging community like LJ.  In LJ, the
communities are declared by the users, and users change interest very
often, creating and disintegrating communities quite often.




\subsection{General features in the dynamics of followers}
\label{universal}

\begin{figure*}
\centering 
\includegraphics[width=0.4\columnwidth]{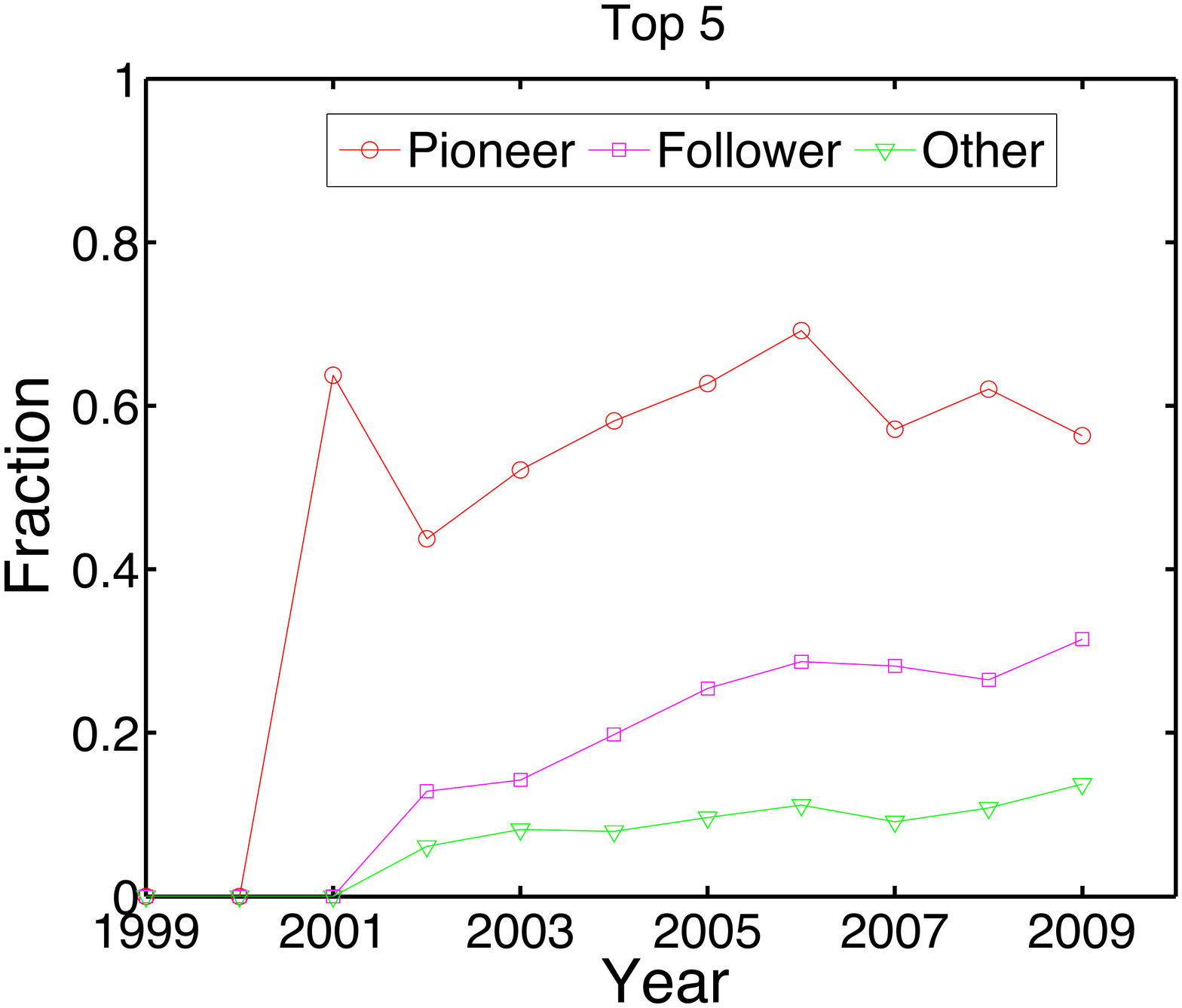}\includegraphics[width=0.4\columnwidth]{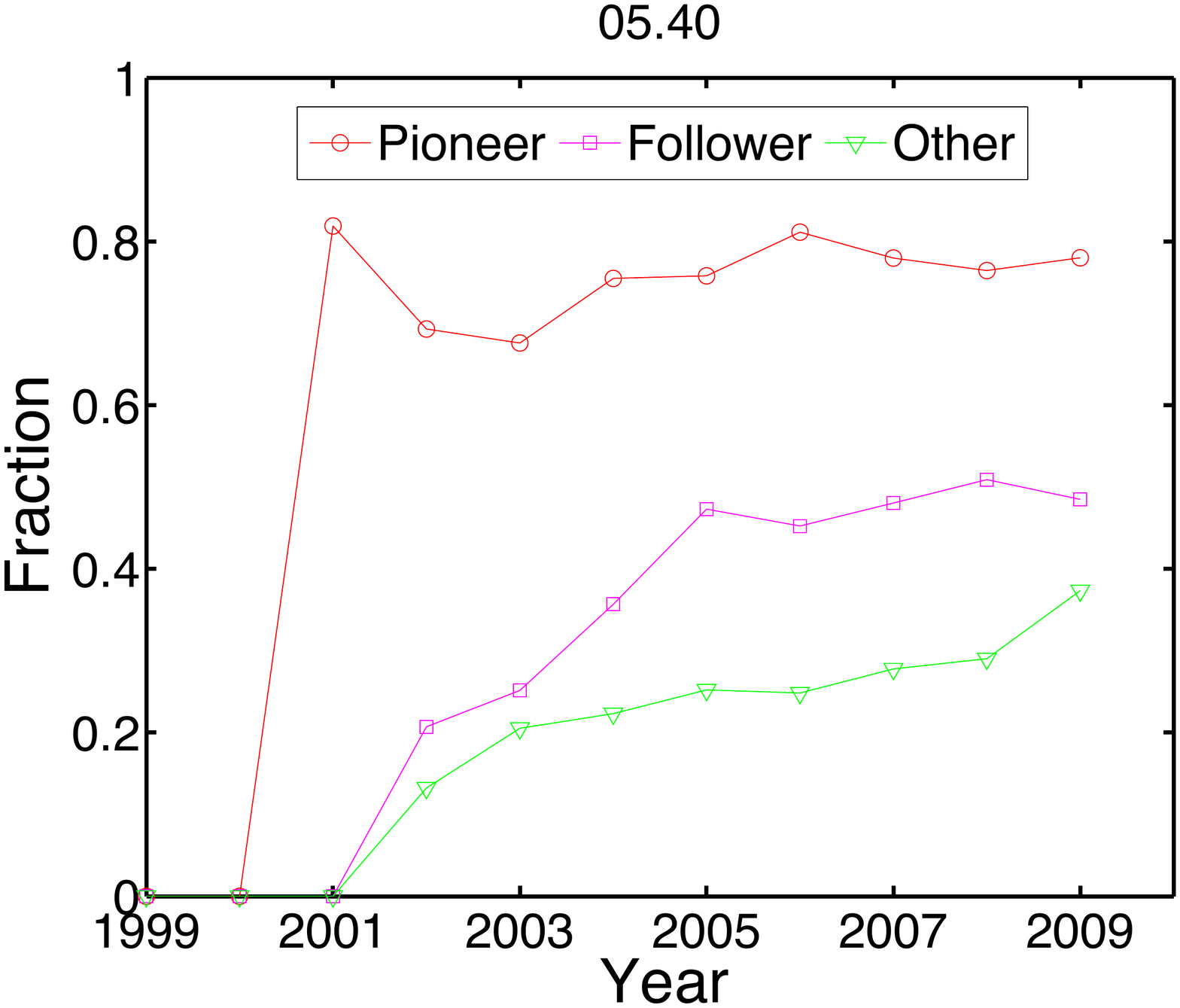}
\includegraphics[width=0.4\columnwidth]{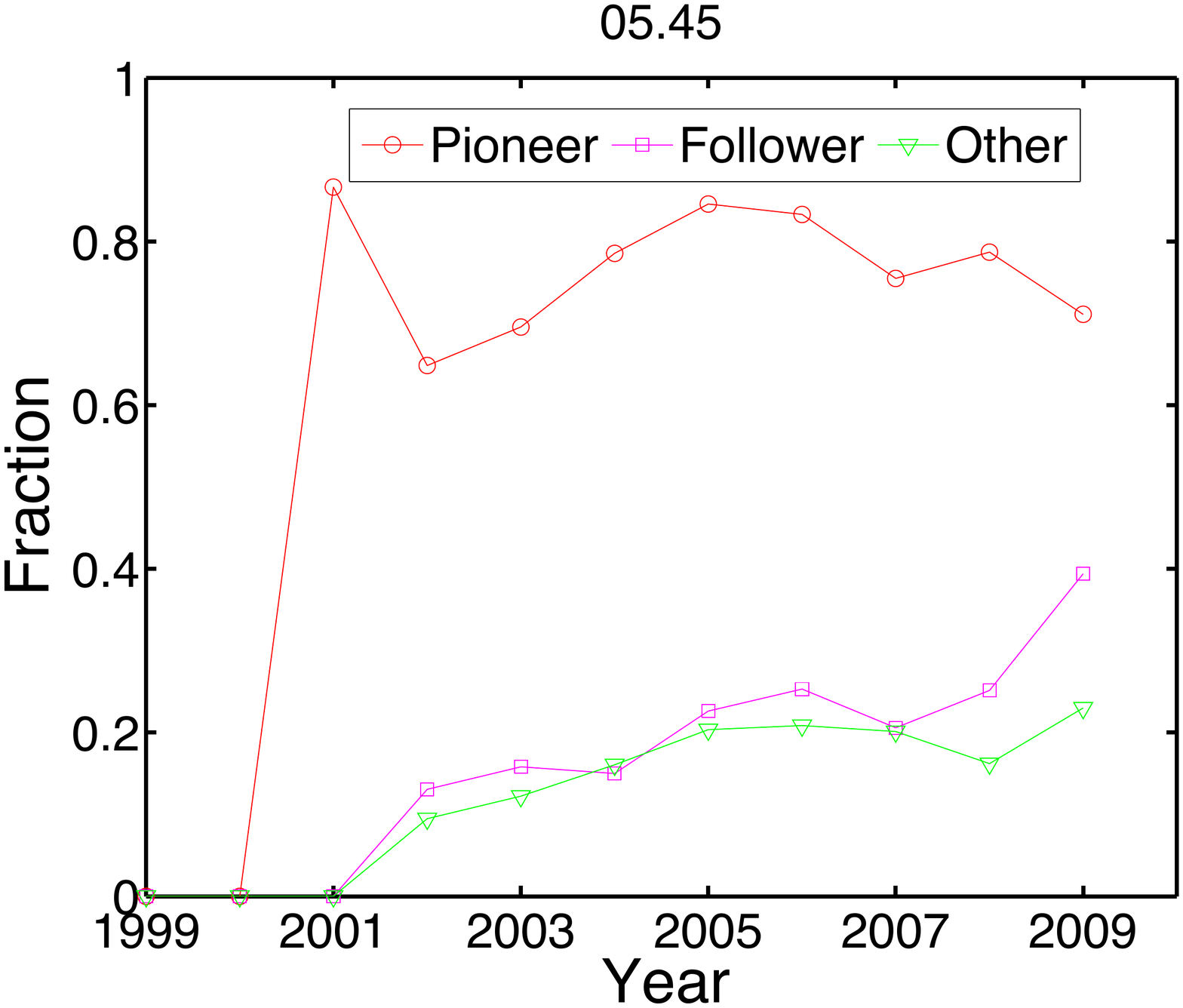}\includegraphics[width=0.4\columnwidth]{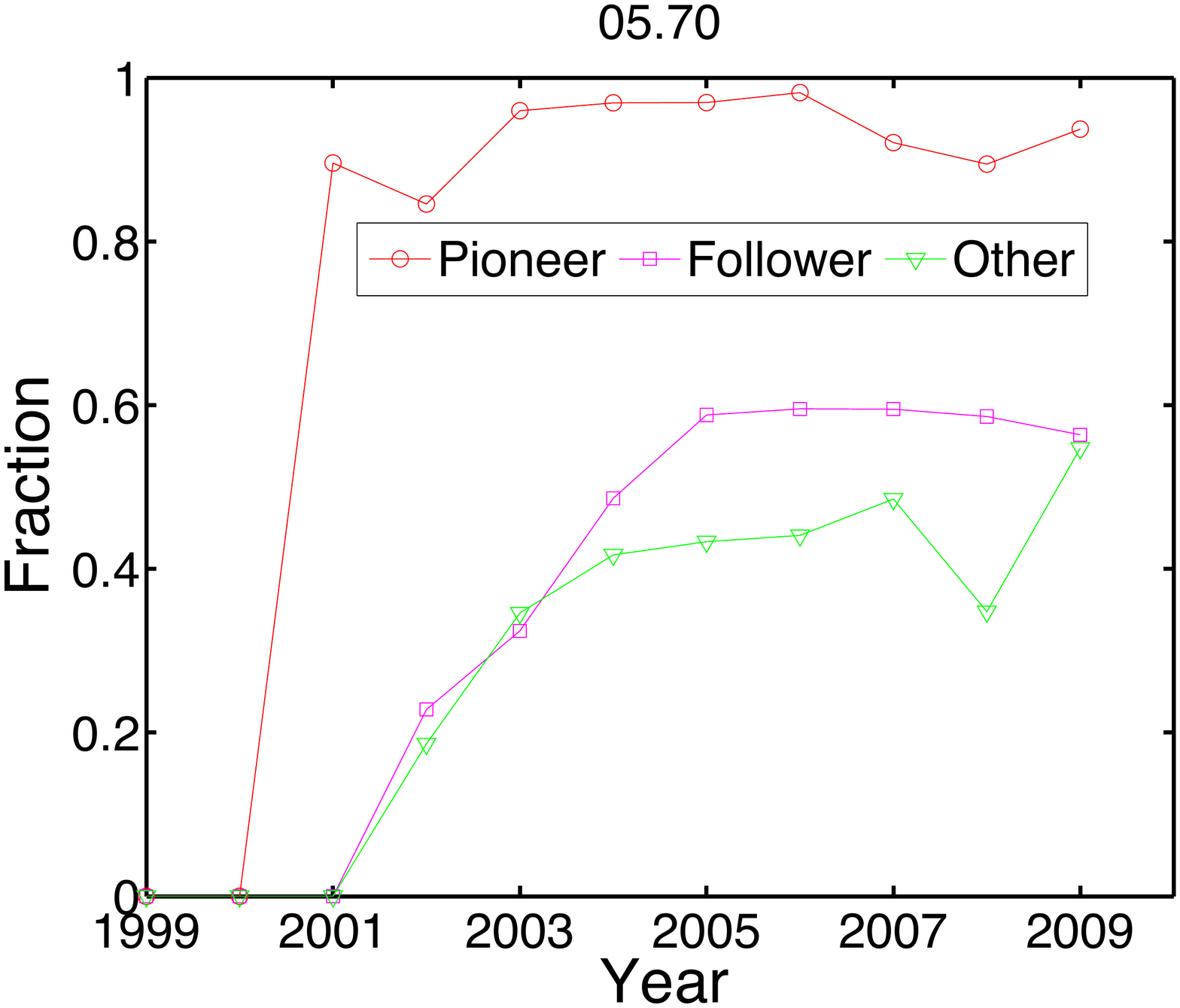}
\includegraphics[width=0.4\columnwidth]{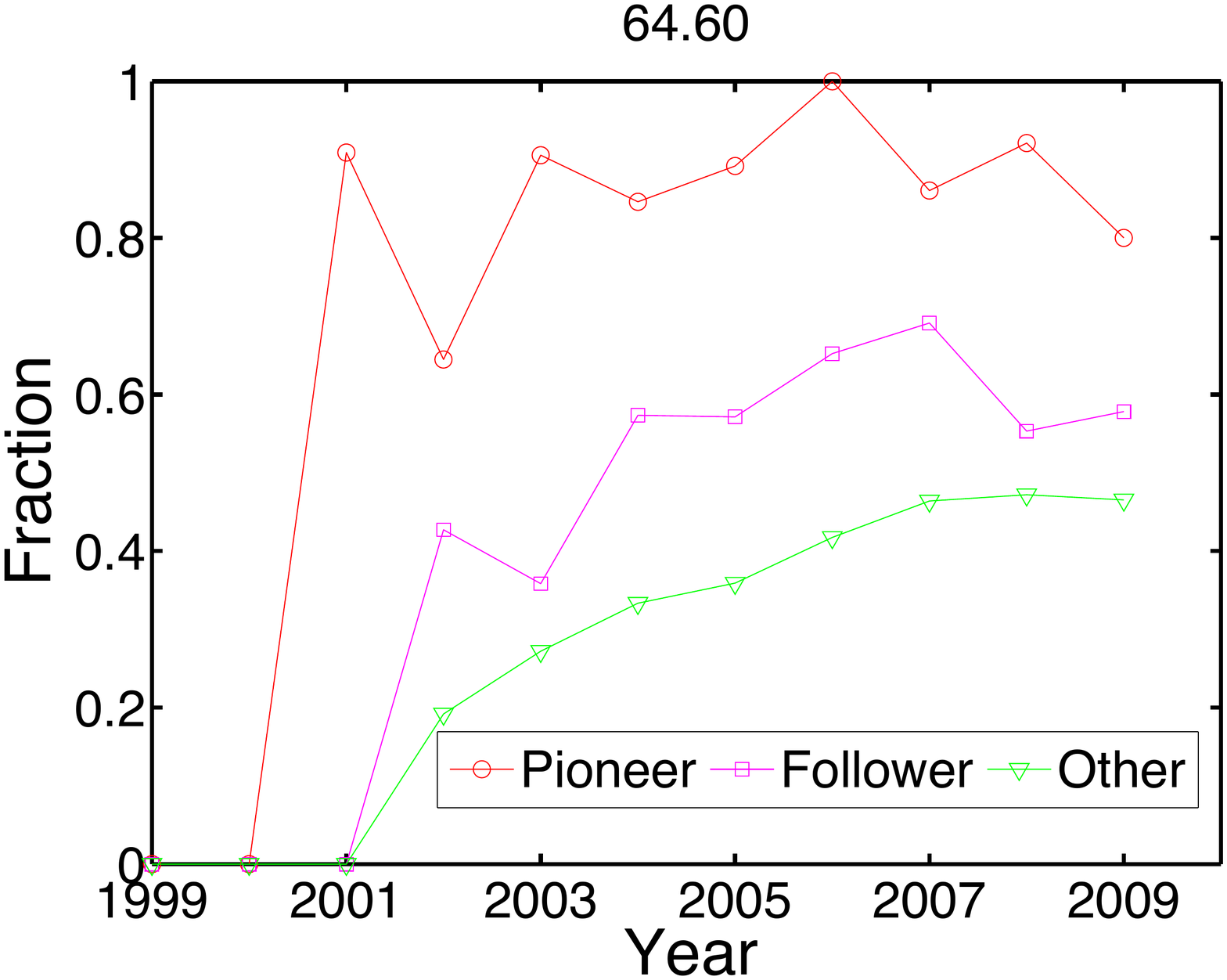}\includegraphics[width=0.4\columnwidth]{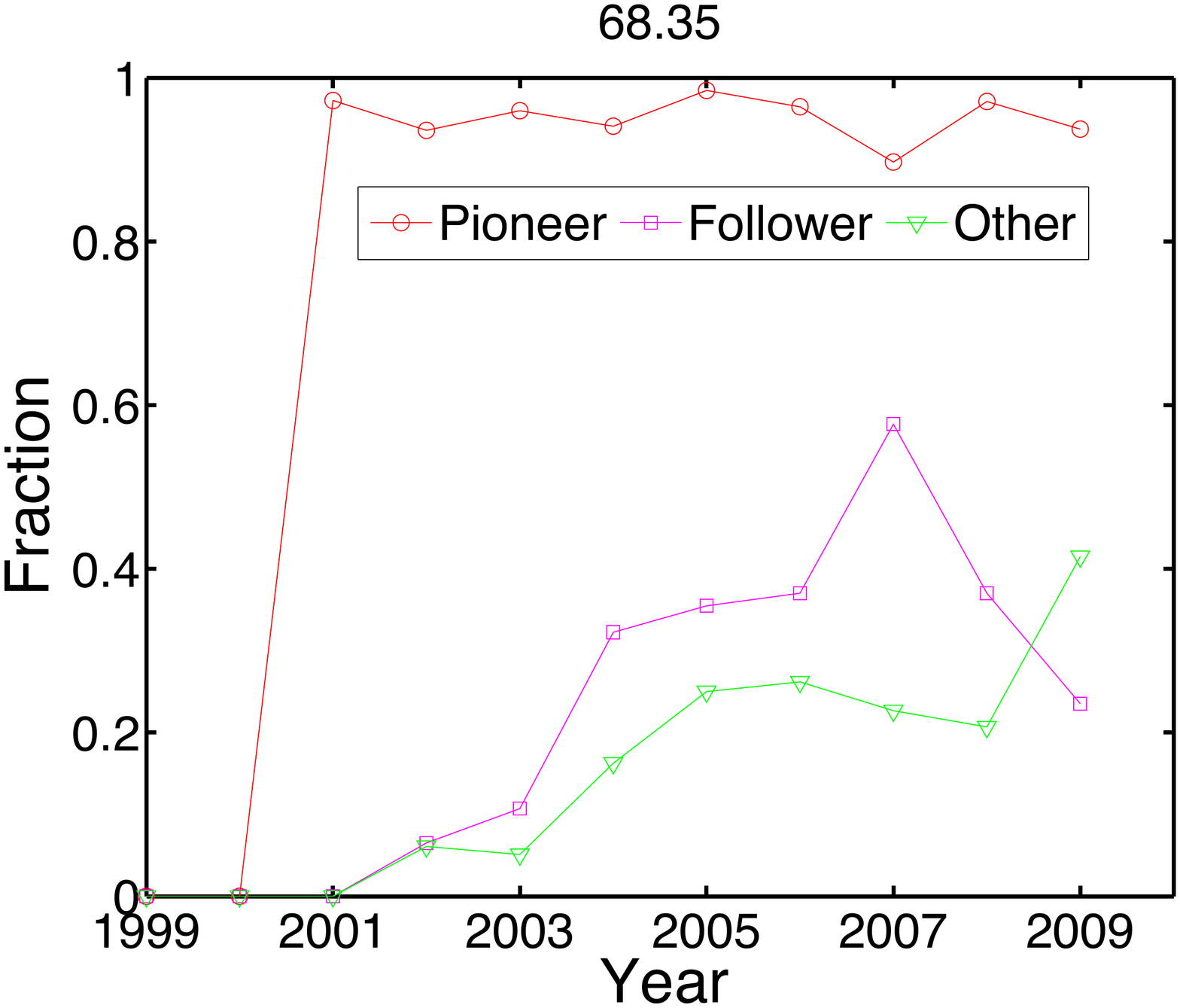}
\caption{ Comparison of the fraction of papers published by
    pioneers, followers and followers of followers. Followers are the
  scientists who at least cited one paper of the pioneers between 1997
  to 2001 and are not the cooperators of pioneers.  For a given PACS
  number as indicated in the figures, we record the number of papers
  published in the PACS number and the number of papers published only
  in 89.75 (Complex Networks) for each author. Then we get the total
  number of papers published in complex networks $N_{\rm net}$ and the
  number of papers with these two PACS numbers $N_{\rm all}$ for every
  author. The vertical axis shows the fraction $\frac{N_{\rm
      net}}{N_{\rm all}}$ as a function of time, which approximately
  satisfies pioneers$>$followers$>$the rest. This result suggests that
  influence links are important for cascading dynamics.
\label{PioF1Other}}
\end{figure*}

Evidence of the cascading processes triggered by the pioneers is
depicted in SI Fig. \ref{PioF1Other} showing general
features in the dynamics of followers. We track the pioneers who first
published in Complex Networks in 2001, then the followers and then the
rest. For each of the top 5 PACS number with 89.75 together, we
calculate the largest component of the coauthorship network according to
the APS publication records from 1997 to 2001. Using this data, we
detect the pioneers of network science and all of the followers of
pioneers in this coauthorship network. SI Fig. \ref{PioF1Other} shows
general features in the dynamics of followers. We compare the fraction
of papers published in complex networks among pioneers, followers and
the rest for different PACS number form 2001 to 2009 and found that
the fraction satisfies that pioneers$>$followers$>$the rest. This
suggests that influence links follows the influence links in a cascade
of followers.

\section{Generating function theory of percolation in correlated influence
networks}
\label{recursive}

We consider a network with both, bidirectional connectivity links and
directed influence links. Each node has three degrees: $(k, k_{\rm
  in}, k_{\rm out})$ measuring the number of connectivity links,
in-going influence links and out-going influence links, respectively.
The properties of such a network are described by the three
dimensional generating function:
 \begin{equation}
G(x_1,x_2,x_3)=\sum_{k,k_{\rm in},k_{\rm out}}P(k,k_{\rm in},k_{\rm
  out})x_1^kx_2^{k_{\rm in}}x_3^{k_{\rm out}},
\end{equation}
where the joint probability distribution $P(k,k_{\rm in},k_{\rm out})$
describes the local correlations among $(k, k_{\rm in}, k_{\rm
  out})$.  In the following we denote the higher order derivatives as:
\begin{equation}
G_{x_1^{m_1},x_2^{m_2},x_3^{m_3}}(x_1,x_2,x_3)=\frac{\partial^{m_1+m_2+m_3}G(x_1,x_2,x_3)}{\partial x_1^{m_1}\partial x_2^{m_2}\partial x_3^{m_3}}.
\end{equation}

If node $i$ is being influenced by node $j$, there is a directed
influence link from node $i$ to node $j$.  In a real system, the
influence of the piers is applied with a given probability $q_h$ which
is less than 1. That is, even if there is an influence link from $i$
to $j$, if $j$ departs, then $i$ will depart with probability $q_h$
which in general is smaller than 1. In SI Section
\ref{influence_links} we explain in detail how to estimate $q_h$ from
the APS data. This effect is taken into account in the model. In order
to simplify the problem, the removal following influence links is
analogous to randomly deleting a $1-q_h$ fraction of influence links,
and then assuming that all of the remaining nodes connected via
influence links are removed with probability 1. Without losing any
generality, in our analysis we set this probability to be
$q_h=1$, as done in previous percolation studies \cite{NewmanSpread}. 
Next, we analyze the cascading process following the recursive stages:
percolation process $\to$ influence-induced percolation process $\to$
percolation process $\to$ influence-induced percolation $\cdots$.

\subsection {First Stage: Classical Percolation Process}

The cascading process is triggered by initially randomly removing a
fraction of $1-p$ nodes.  We use $\tilde{p}$ to denote the fraction of
remaining nodes after initial random removing. Thus, at this first
stage, we have:
\begin{equation}
 \tilde{p}=p.
\end{equation}
The generating function of the connectivity degree distribution
related to the branching process is
$\frac{G_{x_1}(x_1,1,1)}{G_{x_1}(1,1,1)}$\cite{Callaway,PerLong,sergey}. Thus
the giant component size $x$ after the first stage of the percolation
process can be written as:
\begin{equation}
 x=p[1-G(t,1,1)],
\label{s1}
\end{equation}
where,
\begin{equation}
 t=1-\tilde{p}(1-f),
\label{s2}
\end{equation}
and
\begin{equation}
 f=\frac{G_{x_1}(1-\tilde{p}(1-f),1,1)}{G_{x_1}(1,1,1)}.
\label{s3}
\end{equation}
The physical meaning of the quantity $t$ is that, a node with
connectivity degree $k$ will have $1-t^k$ probability staying in the
giant component \cite{SgPRE}. Accordingly, we get the average
in-degree in the giant component of size $x$, $\langle k_{\rm
  in}^x\rangle$, as:
\begin{equation}
\langle k_{\rm in}^x
\rangle=\frac{G_{x_2}(1,1,1)-G_{x_2}(t,1,1)}{1-G(t,1,1)}.
\end{equation}

\subsection{Second Stage: Influence-Induced Percolation  Process}

In order to treat the local correlations between $(k, k_{\rm
  in}, k_{\rm out})$, we develop a generating function theory for the
first time by combining the percolation process on the connective
links and the influence directed links in a correlated fashion.  It is
instructed to treat first the second stage assuming that the network
has influence links but they are uncorrelated and then we will
generalize the results to the existence of correlation.

Let $H(u)$ be the generating function for the probability of reaching
an outgoing component of a given size by following a directed links on
the original network. According to reference \cite{cohen}, $H(u)$ can
be written as
\begin{equation}
\label{h1}
H(u)=u\frac{G_{x_2}(1,1,H(u))}{\langle k_{\rm in} \rangle}.
\end{equation}
Let,
\begin{equation}
H(u)=\sum_s\eta(s)u^s.
\end{equation}

Assuming we can reach $s$ nodes following an out-going directed link
(influence links), thus for any given random subnetwork of size $u$
(randomly selected from the original network), the probability for all
of the $s$ nodes to be in $u$ is $u^s$. This means that the generating
function for the probability that all the nodes reached by following a
directed link are in $u$ is $H(u)$. 

Next, we generalize the above argument by considering a correlated
selection of nodes in $u$, rather than considering a random
uncorrelated selection of nodes in $u$ as above. If we choose the
nodes in a correlated fashion with the in-degree $k_{\rm in}$ only,
any local structure in $u$ are still tree like. Thus the probability
that all the nodes reached by following a directed link are in $u$
is proportional to the sum of the in-degree of all nodes in $u$. Let

\begin{equation}
\hat{u}=\frac{\langle k_{\rm in}^u \rangle}{\langle k_{\rm in}
  \rangle},
\end{equation}
then, the generating function of the probability that all the nodes
reached by following a directed link are in $u$ is
$H(\hat{u})$. Furthermore, a node with $k_{\rm out}$ degree in $u$
will survive with probability $H^ {k_{\rm out}}(\hat{u})$.

Considering the directed network composed by influence links, the
generating function of the out-degree distribution is $G(1,1,x_3)$ and
the corresponding generating function of the out-degree related to the
directed branching process is $\frac{G_{x_2}(1,1,H(x))}{\langle k_{\rm
    in} \rangle}.$ Using $y$ to denote the probability that all the
nodes reached by following a directed link are in $x$, we have
\begin{equation}
y=H(hx),
\label{s4}
\end{equation}
where,
 \begin{equation}
 H(x)=x\frac{G_{x_2}(1,1,H(x))}{\langle k_{\rm in} \rangle},
\label{s5}
\end{equation}
and
\begin{equation}
 h=\frac{\langle k_{\rm in}^x \rangle}{\langle k_{\rm in} \rangle}.
\label{s6}
\end{equation}

\subsection{Third Stage: Recursive Percolation Process}

Removing a node with $k$ and $k_{\rm out}$ in the second stage is
analogous to removing the node with probability $1-py^{k_{\rm out}}$
from the original network in the first stage.  This crucial property
implies that we can map out the cascading process after the second
stage into a new initial removing occurring in the original network.
The fraction of remaining links after the new removing process is:
\begin{equation}
\label{pt1}
 \tilde{p}=\frac{pG_{x_1}(1,1,y)}{G_{x_1}(1,1,1)}.
\end{equation}
Thus, the resulting percolation equations are
\begin{align}\label{mainsys}
\\\nonumber
f=\frac{G_{x_1}[1- \tilde{p}(1-f),1,y]}{G_{x_1}(1,1,y)},
\\\nonumber
t=1- \tilde{p}(1-f),
\\\nonumber
x=p[1-G(t,1,1)].
\end{align}

\subsection{Recursive relations}

The above analysis leads to a set of recurring relations defining the
cascading process. After the second stage, the current cascading
effect can be mapped to a removing process in the original
network. This property allows us to write down the cascading process
as recursive equations. Integrating the above three stages, we can
rewrite the cascading process as following:
\begin{align}\label{mainsys5}
x=p[1-G(1-\tilde{p}(1-f),1,1)],
\\\nonumber
f=\frac{G{x_1}(1-\tilde{p}(1-f),1,y)}{G_{x_1}(1,1,y)},
\\\nonumber
\tilde{p}=\frac{pG_{x_1}(1,1,y)}{G_{x_1}(1,1,1)},
\\\nonumber
t=1-\tilde{p}(1-f),
\\\nonumber
\langle k_{\rm in}^x \rangle=\frac{G_{x_2}(1,1,1)-G_{x_2}(t,1,1)}{1-G(t,1,1)},
\\\nonumber
h=\frac{\langle k_{\rm in}^x \rangle}{\langle k_{\rm in} \rangle},
\\\nonumber
y=hx\frac{G_{x_2}(1,1,y)}{\langle k_{\rm in} \rangle}.
\end{align}

The physical meaning of these recursion relations is that, after each
first stage, a node that is not removed in the initial attack has
$1-t^k$ survival probability. After the second stage, the survival
node can be mapped out to a removing process occurring on the original
network with probability $1-py^{k_{\rm out}}$. These two properties
allow us to write down the formula of the relative size of giant
component $P_{\infty}$ at the final stable state of the cascading
process as:
\begin{equation}
P_{\infty}(p)=p[G(1,1,y)-G(t,1,y)].
\label{pinfty}
\end{equation}

\section{The critical threshold and universal boundary for phase transitions}
\label{critical}

It is of interest to understand the universal properties at the
critical point. Below we derive the critical threshold for first-order
and second-order transitions, and the boundaries between these two
phases. The equations are valid for any type of correlations. In the
following sections we apply our results to uncorrelated and correlated
networks with influence links.

\subsection{Second-order phase transition}
\label{critical_a}

When the transition is of second order, the giant component
$P_{\infty}$ tends to 0 continuously as $p$ approaches the critical
threshold $p_c^{\rm II}$. This implies that when $p\rightarrow
p_c^{\rm II}$, $x \rightarrow 0$ and $y\rightarrow 0$, as well as $t
\rightarrow 1$ and $f \rightarrow 1$, continuously. Let
\begin{equation}
z=\tilde{p}(1-f),
\label{z}
\end{equation}
and the second equation of the main recursive Eqs. (\ref{mainsys5})
can be written as
\begin{equation}\label{1}
    f=\frac{\sum_{k,k_{\rm in},k_{\rm out}}P(k,k_{\rm in},k_{\rm out})k(1-z)^{k-1}y^{k_{\rm out}}}{\sum_{k,k_{\rm in},k_{\rm out}}P(k,k_{\rm in},k_{\rm out})ky^{k_{\rm out}}}.
\end{equation}
Thus,
\begin{equation}\label{2}
  f=\frac{\sum_{k,k_{\rm in},k_{\rm out}}P(k,k_{\rm in},k_{\rm
      out})k\sum_{\mu=0}^{k-1}C^{k-1}_{\mu}(-z)^{\mu}y^{k_{\rm
        out}}}{\sum_{k,k_{\rm in},k_{\rm out}}P(k,k_{\rm in},k_{\rm
      out})ky^{k_{\rm out}}},
\end{equation}

When $p\rightarrow p_c^{\rm II}$, then $z \rightarrow 0$. Therefore,
close to the critical point we can ignore the terms of order $O(z^2)$
and $O(y)$, and the above equation can be written as:

\begin{footnotesize}
\begin{equation}\label{4}
    f=\frac{\sum_{k,k_{\rm in},0}P(k,k_{\rm
        in},0)[k-k(k-1)z+\frac{1}{2}k(k-1)(k-2)z^2]+\sum_{k,k_{\rm
          in},1}P(k,k_{\rm
        in},1)[ky-k(k-1)zy+\frac{1}{2}k(k-1)(k-2)z^2y]}{\sum_{k,k_{\rm
          in},0}P(k,k_{\rm in},0)k+\sum_{k,k_{\rm in},1}P(k,k_{\rm
        in},1)ky}.
\end{equation}
\end{footnotesize}
Using Eq. (\ref{z}), Eq.  (\ref{4}) can be reduced to:
\begin{equation}\label{5}
    z=\frac{\sum_{k,k_{\rm in},0}P(k,k_{\rm
        in},0)[k(k-1)-\frac{1}{\tilde{p}}k]+\sum_{k,k_{\rm
          in},1}P(k,k_{\rm
        in},1)[k(k-1)y-\frac{1}{\tilde{p}}ky]}{\sum_{k,k_{\rm
          in},0}P(k,k_{\rm in},0)\frac{1}{2}k(k-1)(k-2)+\sum_{k,k_{\rm
          in},1}P(k,k_{\rm in},1)\frac{1}{2}k(k-1)(k-2)y}.
\end{equation}
 Using the forth equation in the main system Eqs. (\ref{mainsys5}), we
 obtain:
\begin{equation}\label{6}
    t=t(y)=1-\frac{\sum_{k,k_{\rm in},0}P(k,k_{\rm
        in},0)[k(k-1)-\frac{1}{\tilde{p}}k]+\sum_{k,k_{\rm
          in},1}P(k,k_{\rm
        in},1)[k(k-1)y-\frac{1}{\tilde{p}}ky]}{\sum_{k,k_{\rm
          in},0}P(k,k_{\rm in},0)\frac{1}{2}k(k-1)(k-2)+\sum_{k,k_{\rm
          in},1}P(k,k_{\rm in},1)\frac{1}{2}k(k-1)(k-2)y}.
\end{equation}
Note that, here, $t$ is written as a function of $y$.

When $y\to 0$, the 7th equation of the main system
Eqs. (\ref{mainsys5}) can be written as
\begin{equation}\label{66}
y=y(t)=hx\frac{\sum_{k,k_{\rm in},0}P(k,k_{\rm in},0)k_{\rm in}}{\langle
  k_{\rm in} \rangle}.
\end{equation}

Substituting Eqs. (\ref{mainsys5}) into Eq. (\ref{66}), we obtain:
\begin{equation}\label{7}
    y(t)=\frac{G_{x_2}(1,1,1)-G_{x_2}(t,1,1)}{\langle k_{\rm in}
      \rangle[1-G(t,1,1)]} \cdot
    \frac{p[1-G(t,1,1)]\frac{1}{\langle k_{\rm in}
        \rangle}\sum_{k,k_{\rm in},0}P(k,k_{\rm in},0)k_{\rm
        in}}{1-\sum_{k,k_{\rm in},1}P(k,k_{\rm in},1)k_{\rm in}}.
\end{equation}
Note that, here we write $y$ as a function of $t$.

By simplifying the third equation of the main system (\ref{mainsys5}),
we obtain:
\begin{equation}\label{8}
\tilde{p}=\frac{p\sum_{0,k_{\rm in},k_{\rm out}}p(0,k_{\rm in},k_{\rm
    out})k}{\langle k \rangle}.
\end{equation}
Substituting Eqs.  (\ref{6}) and (\ref{8}) into Eq.  (\ref{7}) and
ignoring terms of higher order in $y$, we obtain the explicit formula
for the critical point in the second order phase transition:
\begin{equation}\label{9}
p_c^{\rm II}=\frac{\langle k \rangle}{G_{x_1^2}(1,1,0)}.
\end{equation}

This equation generalizes the classical uncorrelated percolation
result for networks without influence links, $p_c=\frac{\langle k
  \rangle}{\langle k (k-1) \rangle}$
\cite{cohena,satorras01,barabasi-attack,NewmanSpread,Callaway} to an
influence network with generic correlations, $P(k, k_{\rm in}, k_{\rm
  out})$.

\subsection{First-order phase transition}
\label{critical_b}

\begin{figure*}
\centering
\includegraphics[width=0.6\columnwidth]{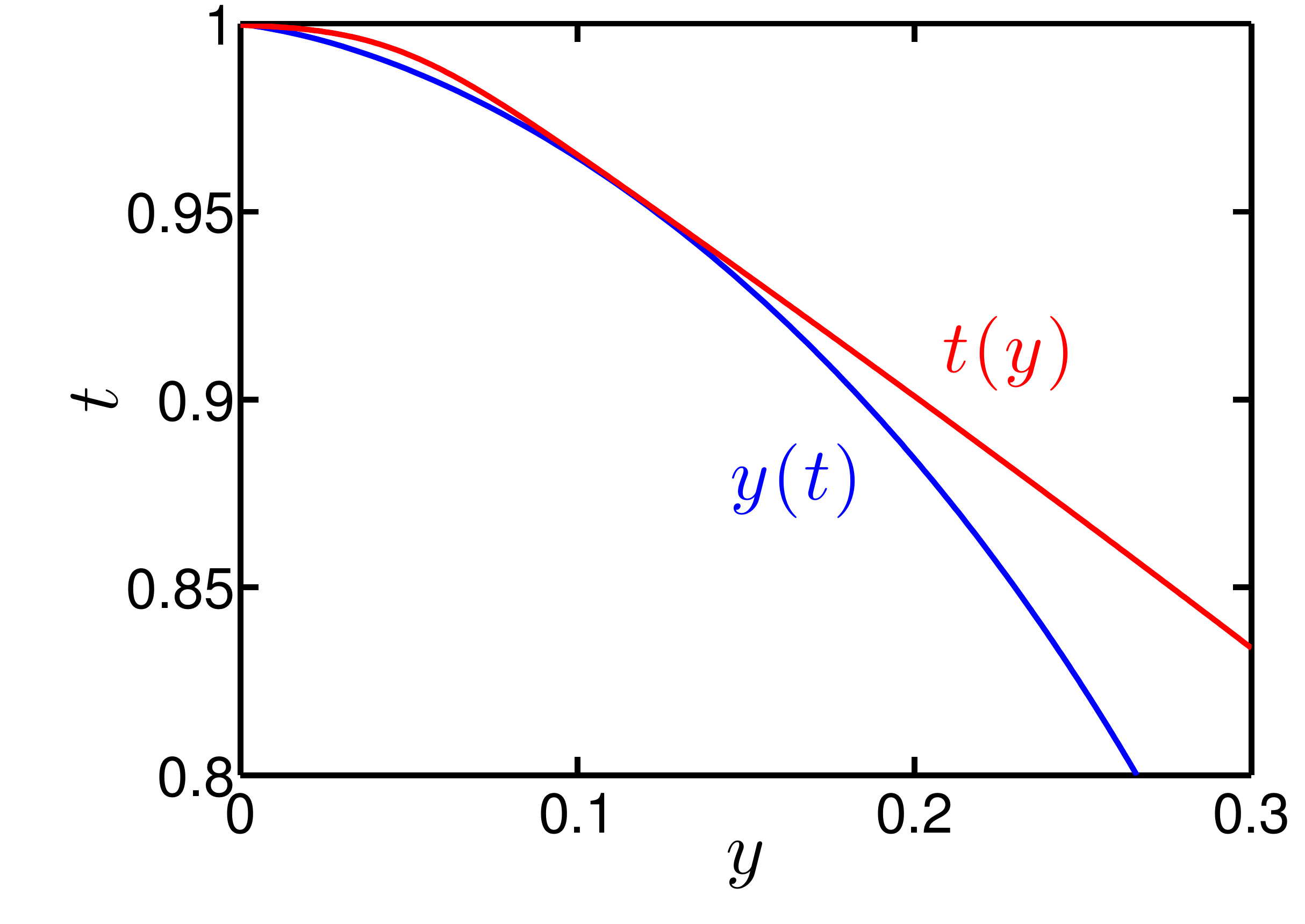}
\caption{ Condition for a first-order phase transition.  This
  figure shows that the functions $t(y)$ and $y(t)$ are tangential at
  the critical point when the transition is of the first order. Here
  $\langle k_{\rm out} \rangle=0.74$, $\gamma=3$ and minimum degree
  $m=2$.}
\label{tangentialattachedfig}
\end{figure*}

For a given $G(x_1,x_2,x_3)$, if there exist a first-order phase
transition, the critical point function $t(y)$, Eq. (\ref{6}) and
$y(t)$ Eq. (\ref{7}) must be tangential to each other (as shown in
SI Fig. \ref{tangentialattachedfig}). Thus, the condition for a
first-order phase transition is
\begin{equation}\label{10}
    \frac{\partial t(y,p_c^{\rm I})}{\partial y}\big |_{y=0} \cdot
    \frac{\partial y(t,p_c^{\rm I})}{\partial t} \big |_{t=1}=1.
\end{equation}
Contrary to the case of $p_c^{\rm II}$ in Eq. (\ref{9}), it is usually
not possible to find an explicit formula for $p_c^{\rm I}$ from
Eq. (\ref{10}). Therefore, we resort to a numerical integration of
Eq. (\ref{10}).

\subsection{Boundary between transitions}

According to Eq.  (\ref{6}), (\ref{8}) and (\ref{9}), we obtain:
\begin{equation}\label{11}
\frac{\partial t}{\partial y}\big |_{y=0}=-\frac{\sum_{k,k_{\rm
      in},1}P(k,k_{\rm
    in},1)\Big[k(k-1)-\frac{G_{x_1}(1,1,1)}{p_c^{\rm
        I}G_{x_1}(1,1,0)}k\Big]-\frac{p_c^{\rm
      I}G_{x_1,x_3}(1,1,0)G_{x_1}(1,1,1)}{[p_c^{\rm
        I}G_{x_1}(1,1,0)]^2}\sum_{k,k_{\rm in},0}P(k,k_{\rm in},0)k
}{\frac{1}{2}\sum_{k,k_{\rm in},0}P(k,k_{\rm in},0)k(k-1)(k-2)},
\end{equation}
and
\begin{equation}\label{12}
\frac{\partial y}{\partial t}\big |_{t=1}
=\frac{-G_{x_1,x_2}(1,1,1)p_c^{\rm I}G_{x_2}(1,1,0)}{\langle k_{\rm
    in} \rangle^2}.
\end{equation}
Substituting Eqs. (\ref{11}) and (\ref{12}) into Eq. (\ref{10}) we
obtain the boundary between the first and second order phase
transitions regime in the phase space:
\begin{equation}\label{13}
1=\frac{2p_c^{\rm II}\langle k \rangle G_{x_1^2,x_3}(1,1,0)hG_{x_2}(1,1,0)}{\langle k_{\rm in} \rangle G_{x_1^3}(1,1,0)}.
\end{equation}

There also exists an additional boundary between a stable network and
unstable network with $p_c=1$. Such a network will disintegrate after
removing a vanishing fraction of nodes in the thermodynamic limit. 

\section{Networks with uncorrelated influence links}
\label{uncorrelated}

We treat the case of a random uncorrelated network with influence
links but no local correlations among $(k, k_{\rm in}, k_{\rm
  out})$. In this case $h\equiv1$, and the main system
(\ref{mainsys5}) can be reduced to:
\begin{eqnarray}\label{uncorsys}
x=p[1-G(1-\tilde{p}(1-f),1,1)],
\\\nonumber
f=\frac{G{x_1}(1-\tilde{p}(1-f),1,y)}{G_{x_1}(1,1,y)},
\\\nonumber
\tilde{p}=\frac{pG_{x_1}(1,1,y)}{G_{x_1}(1,1,1)},
\\\nonumber
t=1-\tilde{p}(1-f),
\\\nonumber
y=x\frac{G_{x_2}(1,1,y)}{\langle k_{\rm in} \rangle}.
\\\nonumber
P_{\infty}=p[G(1,1,y)-G(t,1,y)].
\end{eqnarray}
Simplifying the above system, at the equilibrium state the giant
component $P_{\infty}$ can be written as:
\begin{eqnarray}\label{newnewmain}
P_{\infty}=pG(1,1,P_{\infty})\{1-G[1-pG(1,1,P_{\infty})(1-f),1,1]\},
\end{eqnarray}
and
\begin{eqnarray}
f=\frac{G_x[1-pG(1,1,P_{\infty},1,1)(1-f)]}{G_x(1,1,1)}.
\end{eqnarray}

When the transition is second order, the critical point can be
obtained explicitly as:
\begin{eqnarray}\label{highlights5}
p_c^{\rm II, unc}=\frac{\langle k \rangle}{q_0
  \langle k(k-1) \rangle},
\end{eqnarray}
where, $q_0$ is the fraction of nodes with $k_{out}=0$. Therefore, for
a scale-free network with $\gamma<3$, we find $p_c^{\rm II, unc} = 0$,
since the second moment diverge, despite the presence of the influence
links, as long as these links are uncorrelated with each other.

\section{Networks with correlated influence links}
\label{correlated}

\begin{figure*}[hbt]
\centering
\includegraphics[width=0.5\columnwidth]{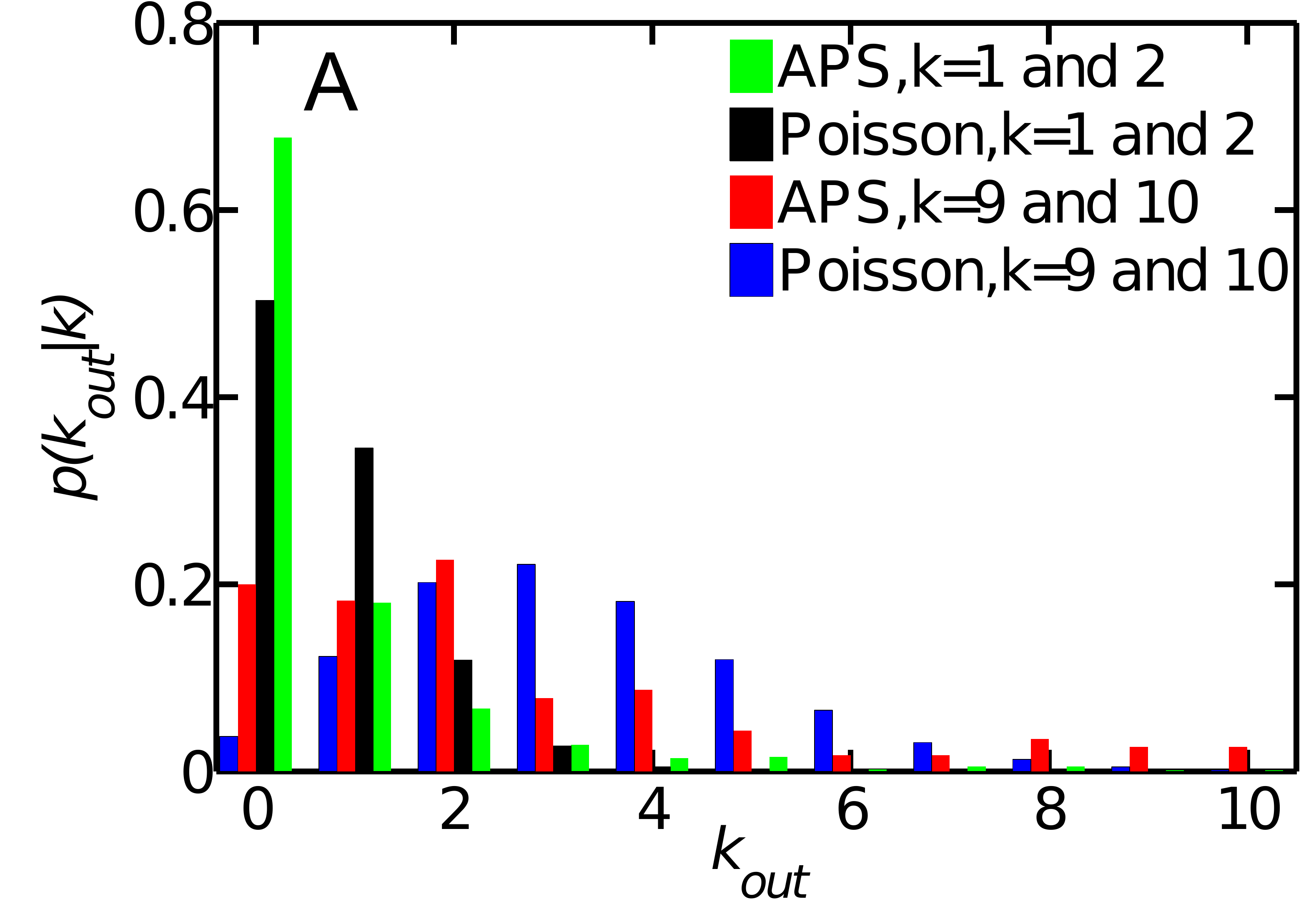}\includegraphics[width=0.5\columnwidth]{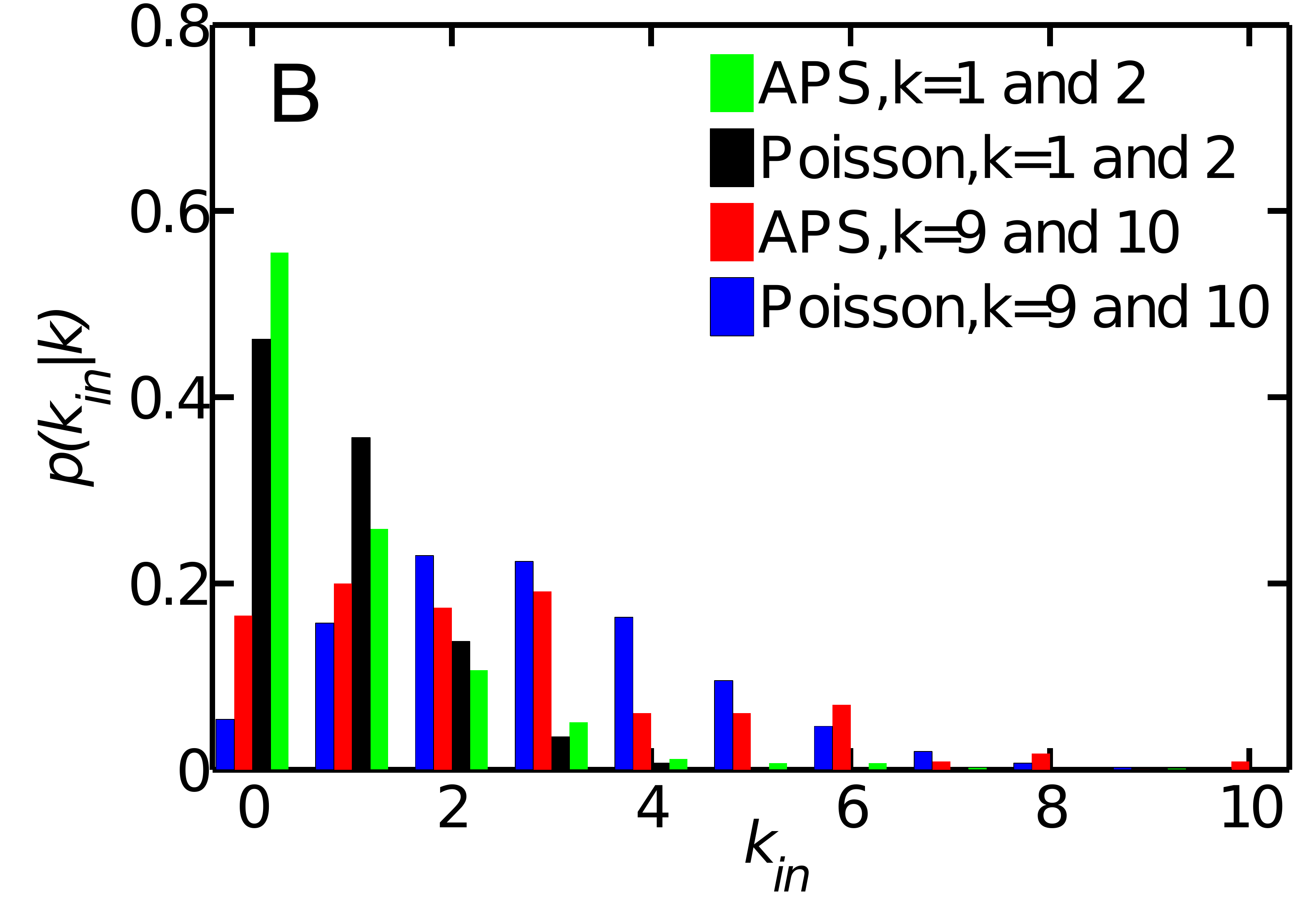}
\caption{ Distribution of (a) $P(k_{\rm out}|k)$ and (b) $P(k_{\rm
      in}|k)$ for the five APS communities. We calculate the
  probability of $k_{\rm in}$ and $k_{\rm out}$ for a given $k$ and
  find that they can be approximated by Poisson distributions.}
\label{poisson}
\end{figure*}

When $k_{in}$ and $k_{out}$ correlate with $k$, the generating
function $G(x_1,x_2,x_3)$ can be written as:
\begin{equation}
G(x_1,x_2,x_3)=\sum_kP_1(k)x_1^k\cdot\sum_{k_{in}}P_2(k_{in}|k)x_2^{k_{in}}\cdot\sum_{k_{out}}P_2(k_{out}|k)x_2^{k_{out}}.
\end{equation}
For a given $P_1(k)$ and $(\alpha, \beta, \langle k_{\rm in}
\rangle)$, the average in-degree and out-degree for given $k$ are
$\frac{k^{\beta}}{\langle k^{\beta} \rangle}\langle k_{\rm in}
\rangle$, where $\langle k^{\alpha} \rangle=\sum_{k} P_1(k)k^{\alpha}$
and $\langle k^{\beta} \rangle=\sum_{k} P_1(k)k^{\beta}$. $P_2(k_{\rm
  in}|k)$, $P_3(k_{\rm out}|k)$ is assumed to be both Poisson
distributions (see SI Fig. \ref{poisson}), thus the generating
functions for $P_2(k_{\rm in}|k)$ and $P_3(k_{\rm out}|k)$ are
$\sum_{k_{in}}P_2(k_{in}|k)x_2^{k_{in}}=e^{\frac{1}{\langle k^{\beta}
    \rangle}\langle k_{\rm in} \rangle k^{\beta}(x_2-1)}$ and
$\sum_{k_{out}}P_2(k_{out}|k)x_2^{k_{out}}=e^{\frac{1}{\langle
    k^{\alpha} \rangle}\langle k_{\rm in} \rangle k^{\alpha}(x_3-1)}$.
Therefore, the main generating function can be written as:
\begin{equation}\label{14a}
G(x_1,x_2,x_3)=\sum_{k}P_1(k)x_1^ke^{\frac{1}{\langle k^{\beta}
\rangle}\langle k_{\rm in} \rangle
k^{\beta}(x_2-1)}e^{\frac{1}{\langle k^{\alpha} \rangle}\langle
k_{\rm in} \rangle k^{\alpha}(x_3-1)}.
\end{equation}

This simple formula allow us to solve for $p_c^{\rm II}$ for any
$\alpha$ and $\beta$. Using Eq. (\ref{14a}), we can write the
critical threshold as:
\begin{equation}\label{14}
p_c^{\rm II,cor}=\frac{\langle k \rangle}{\sum_kk(k-1)P(k)\exp\Big(
  {\frac{-k^{\alpha} \langle k_{\rm in} \rangle }{\langle k^{\alpha}
      \rangle} \Big)}}.
\end{equation}

When there is no local correlation between $k$ and $k_{\rm
  out}$, $\alpha=0$ and $\exp\Big({\frac{-k^\alpha \langle k_{\rm
      out}\rangle}{\langle k^\alpha \rangle}}\Big)=\exp\Big(-\langle
k_{\rm out}\rangle \Big)= q_0$ which is consistent with the
uncorrelated result Eq. (\ref{pc_unco}).

Equation (\ref{14}) indicates that $p_c^{\rm II,cor}$ only depends on
$\alpha$, the connectivity degree distribution $P(k)$ ($\gamma$ in the
case of a scale-free network) and the average in-degree or out-degree
since $\langle k_{\rm in} \rangle = \langle k_{\rm out} \rangle$.
Furthermore, $p_c^{\rm II,cor}$ can be different from zero even for a
scale-free network with $\gamma<3$--- unlike the uncorrelated
counterpart, Eq. (\ref{highlights5})--- since the exponential term
stabilizes the divergence of the second moment when $\alpha \ne 0$.
Surprisingly, there is no $\beta$ dependence, implying that the large
correlation between $k$ and $k_{\rm in}$ plays a secondary role in
sustaining the cascades or in changing $p_c^{\rm II,cor}$. When the transition
is of the first order, $p_c^{\rm I,cor}$ depends on both $\alpha$ and $\beta$.

\section{Estimation of distribution functions based on exponential random graph models}
\label{ergm}

Exponential-family random graph models \cite{ERGMWasserman1} are very
useful statistical tools to capture the essential properties of
networks. Here, we employ this tool to estimate the power law exponent
of the degree distribution. For the exponential-family models, we
define
\begin{equation}\label{mef}
P_\theta(\Omega=\omega) = \exp\Big[\sum_{k=0}^{n}\theta_k
  p_k(\omega)\Big],
\end{equation}
where $p_k(\omega)$ is the probability of randomly choosing a node
with degree $k$ in network $\omega$.  We use conditional logistic
regression method to estimate the full vector $\theta$.  It is a
pseudo-likelihood method
\cite{Hunter} which estimate the
vector $\theta$ by maximizing
\begin{equation}
l(\theta)=\sum_{i,j}\Big[P_\theta\big(\Omega_{i,j}=\omega_{i,j}\big|
\Omega_{u,v}=\omega_{u,v}, (u,v)\neq(i,j)\big)\Big].
\end{equation}

After obtaining the vector $\theta$, we employ the Gibbs sampling to
update the links of the network. For a given initial network
$\omega^{1}$ (step 1), the links of this network are stochastically
updated. In step $t$, the probability of the existence of the link
between node $i$ and $j$ is
\begin{equation}
P_\theta\Big(\omega_{i,j}^{t+1}=1\big|\omega_{u,v}^{t+1}=\omega_{u,v}^{t},
(u,v)\neq(i,j)\Big).
\end{equation}

The distribution of the network $\omega^{t}$ converges for $t\to
\infty$ to the exponential random graph distribution.  The above
updating process is so-called Markov chain Monte Carlo (MCMC)
\cite{Hunter}. In this way, we can get a lot of network samples
and the corresponding degree sequences. It allows us to estimate the
power law exponent of the original networks by bootstrap method. We
employ the standard maximum likelihood methods \cite{maxlikehood} to
estimated the power law exponent of degree distribution for each
networks and get the average value and standard deviation of this
exponent is $\gamma=2.9\pm 0.01.$

\section{Estimation of  the average number of influence links}
\label{influence_links}

\begin{figure*}
\centering
\includegraphics[width=0.7\columnwidth]{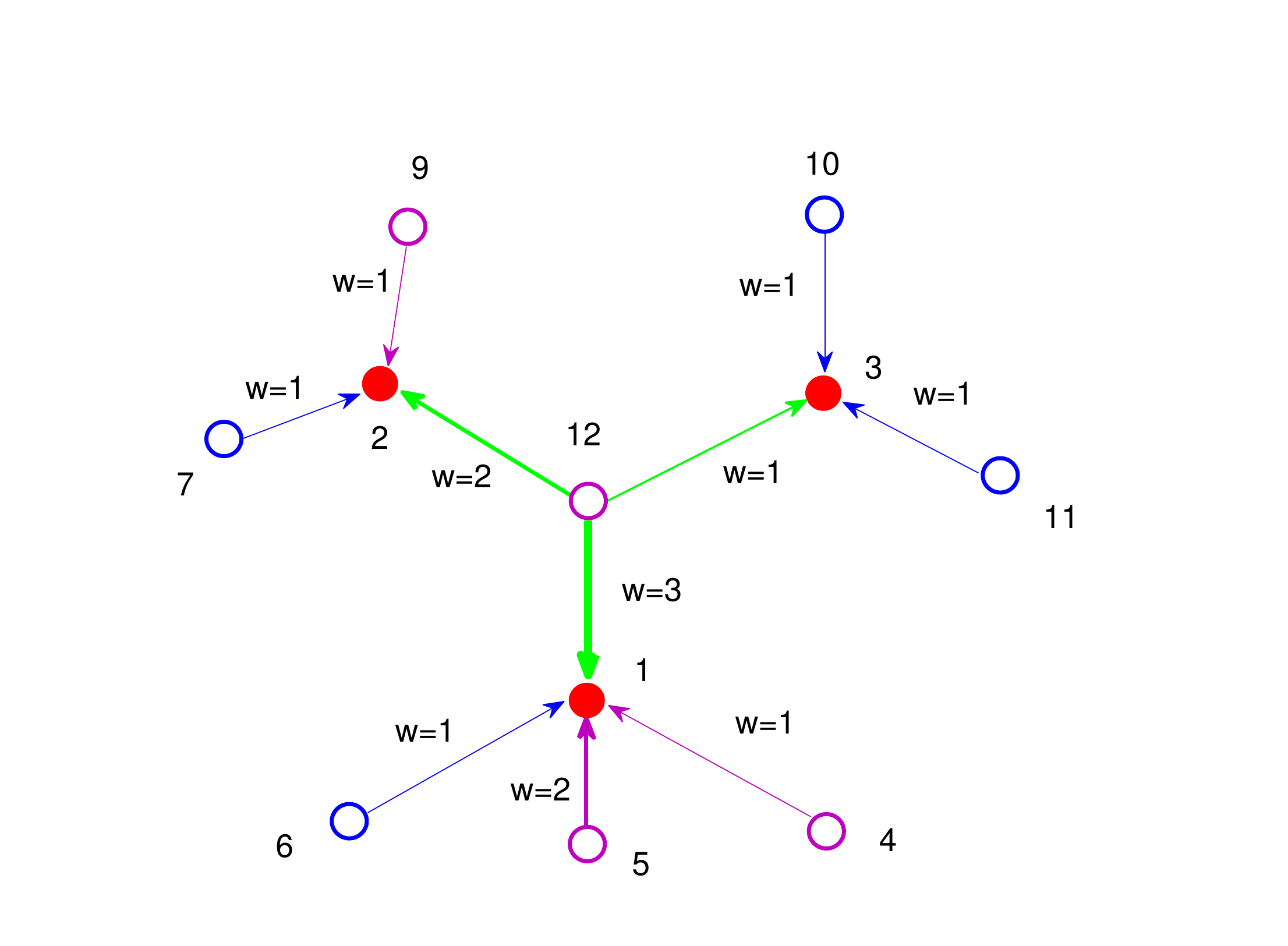}
\caption{Estimation of $f^{\rm min}_w$ and $f^{\rm max}_w$ to obtain
  the bounds in $\langle k_{\rm out} \rangle$ from the empirical data.
  In the figure, we show three pioneers (red nodes), citation links
  with weights $w$ given by the number of citations, and all of the
  followers of the pioneers (open circle). The purple open circles
  denote the followers who move to Complex Networks, while the blue
  open circles do not move.  For the calculating of $f^{max}_w$, we
  consider the links that are active as those from followers who move
  to Complex Networks. In the figure, all the green and purple
  directed links are active. We estimate $f^{\rm max}_w=\frac{a^{\rm
      max}_w}{A_w}$ and $f^{\rm min}_w=\frac{a^{\rm min}_w}{A_w}$ as
  indicated in the text.  For the calculation of $f^{\rm min}_w$, if a
  follower who moves to Complex Networks depends on several pioneers,
  we consider the influence link with the largest weight. For
  instance, for all the green influence links in the figure, the links
  from node 12 to 1 with weight 3 is active and the links from 12 to 2
  and 3 are not active. In this case we obtain: $f^{\rm
    max}_1=\frac{3}{7}, f^{\rm max}_2=\frac{2}{2}=1, f^{\rm max}_3=1$,
  and $f^{\rm min}_1=\frac{2}{7}, f^{\rm min}_2=\frac{1}{2}, f^{\rm
    min}_3=1.$}
\label{akin}
\end{figure*}

First, we detect all the pioneers who have published at least one
Complex Network paper in 2001 and then find all authors who have cited
this pioneers (the followers). Then we identify which followers have
published a paper in Complex Networks in 2002-2003.  We use this
information to estimate, from the real data, the maximum fraction of
influence links for each number of citations $w$, as shown in SI
Fig. \ref{akin}. The lower and upper bounds of the averages $\langle
k_{\rm out} \rangle (=\langle k_{\rm in}\rangle)$ are estimated from
the interval $[ \frac{\sum_w n_wf^{\rm min}_w}{N},\frac{\sum_w
    n_wf^{rm max}_w}{N}],$ where $n_w$ is the number of influence
links with weight $w$ in the giant component, $N$ is the size of the
giant component and $f^{\rm min}_w$, $f^{\rm max}_w$ are the minimum
and maximum active influence links estimated as follows.

To calculate $f^{\rm max}_w$, we consider the number of authors that
cite a given author and calculate the number of influence links which
are active as those from a follower who actually leave the network:
$f^{\rm max}_w=\frac{a^{\rm max}_w}{A_w}$ and $f^{\rm
  min}_w=\frac{a^{\rm min}_w}{A_w}$, where $a^{\rm max}_w$ and $a^{\rm
  min}_w$ are the maximum and minimum number of active influence links
with weight $w$, and $A_w$ is the number of links from the followers
to the pioneers including both active and inactive links with weights
$w$.  For the calculation of $f^{\rm min}_w$, if a follower who moves
to Complex Networks depends on several pioneers, we consider the
influence link with the largest weight.  SI Figure \ref{akin}
illustrates the calculation with an example. Applying this calculation
to the APS communities we find the lower and upper bounds of the
average influence links as $\langle k_{\rm out} \rangle \in
[0.44,0.83]$. The lower bound $\langle k_{\rm out} \rangle = 0.44$ is
used to calculate $P_\infty(p)$ in Fig. \ref{important_figure}a, which
shows how all the empirical data is between the estimated bounds.

To measure the correlations between connectivity degree
and in- and out-degree, first, we employ the above method to measure
the influence probability for each directed influence link for a given
network. Different weights of the directed links gives rise to
different probabilities.  We keep the directed links with these
probability and record all pairs $(k, k_{out})$, $(k, k_{in})$. We
repeat this calculation 20 times and compute the average in- and
out-degree for all the node whose connectivity degree is $k$. Using
the pairs $(k, k_{\rm out})$ and $(k, k_{\rm in})$, the 
correlation scaling law of Fig. \ref{important_figure}d can be
obtained.

\section{Test of correlations between the connectivity and influence networks}

We investigate the correlations between both networks by measuring the
average influence degree (in and out) of the nearest neighbors,
$\langle k_{\rm nn} \rangle$, of a node with connectivity degree $k$
connected via an influence link.

\begin{figure*}
\centering
\includegraphics[width=0.7\columnwidth]{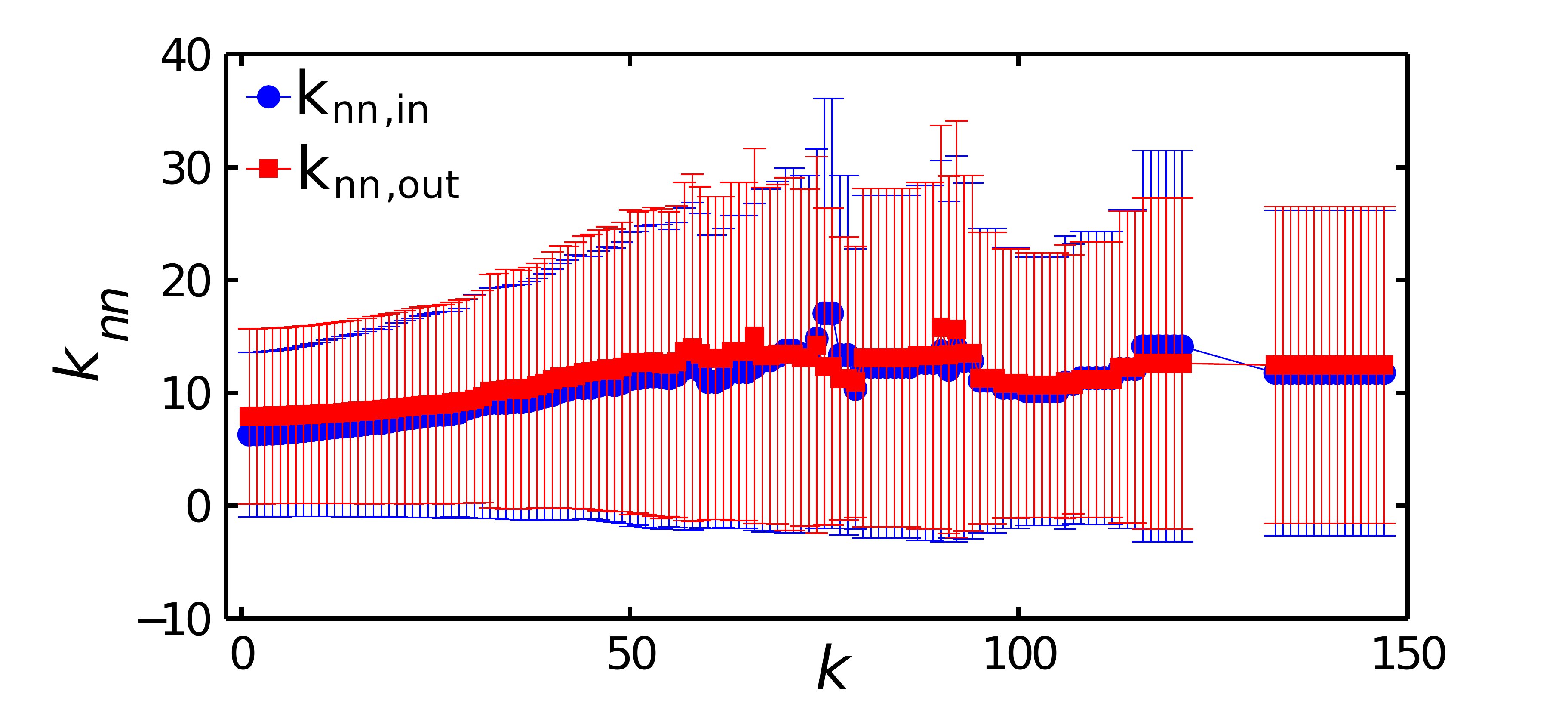}
\caption{ Correlations between connectivity and influence
    networks. A flat line indicates no correlations in the APS
  networks for $\langle k_{\rm nn}\rangle$ for the in-degree and
  $\langle k_{\rm nn}\rangle$ for the out-degree.  In the degree
  sequence, there are no nodes with degree around $k=130$.  }
\label{knn}
\end{figure*}

For the nodes with degree $k$, we measure the average degree $\langle
k_{\rm nn} \rangle$ of their following and follower nodes. We find
that there is almost no correlations between $\langle k_{\rm nn}
\rangle$ and $k$ as shown in Fig. \ref{knn}.  The lack of correlations
is indicated by a flat curve in the plot. We note that due to the fact
that the degree distribution is power-law, the second moment is very
broad, so that it makes the error bars (standard deviation) very
large.

This result implies that the correlation between the
connectivity networks and influence networks themselves are not
significant in the APS, such that the present theory may suffice to
capture the cascading effect. However, the correlations between both
networks might become significant in other networks. In this case, the
theory can be modified to incorporate these type of correlations. In
such a case, the generating function should be generalized to a
six-dimensional function as follows:
\begin{equation}
G(x_1,x_2,x_3, x_4,x_5,x_6)= \sum P(k^1 ,k^1_{in}, k^1_{out}, k^2
,k^2_{in}, k^2_{out}) x_1^{ k^1 }x_2^{ k^1_{in} }x_3^{ k^1_{out} }
x_4^{ k^2 }x_5^{ k^2_{in} }x_6^{ k^2_{out} }
\end{equation} 
to describe the coupled network system. Such a system will not be
difficult to solve with the same techniques developed so far.  The
percolation techniques for a degree-correlated network would be
analogous to those described in the works of Boguna M, Pastor-Satorras
R, Vespignani A. (2003) Absence of epidemic threshold in scale-free
networks with degree correlations. Phys. Rev.  Lett.  {\bf 90}, 028701;
and Goltsev AV, Dorogovtsev SN, Mendes JFF (2008) Percolation on
correlated networks. Phys. Rev. E {\bf 78}, 051105.

\begin{figure*}
\centering
\includegraphics[width=0.7\columnwidth]{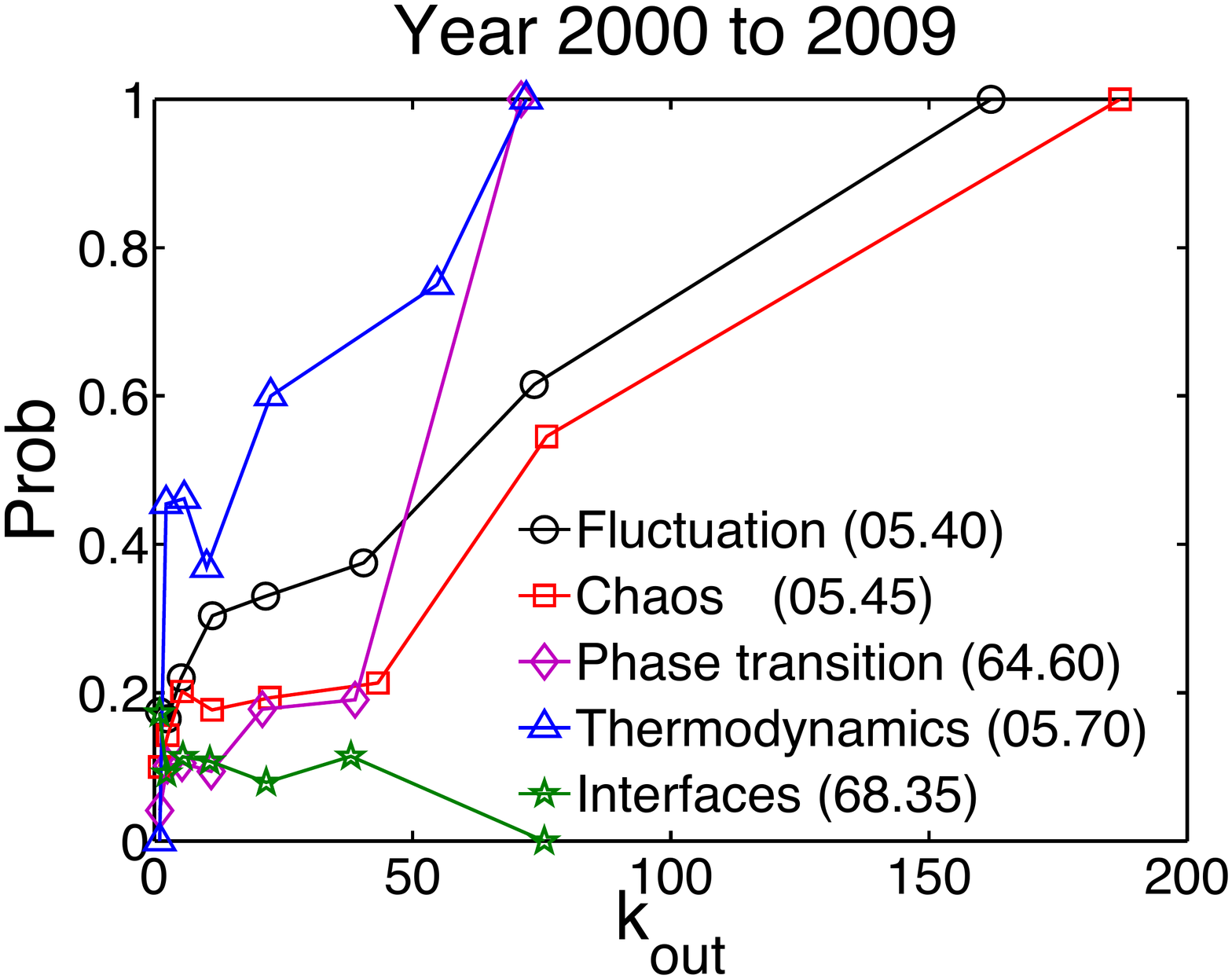}
\caption{ We measured the departure probability of nodes as a function
  of $k$ and $k_{\rm out}$ (number of followee). We observe that the
  hubs with larger number of cooperators $k$ and that are influenced
  by more scientists (larger $k_{\rm out}$) will move into a new field
  with larger probability.}
\label{follow}
\end{figure*}


\end{document}